\documentclass[a4paper,11pt]{article}
\pdfoutput=1 
      
\usepackage{jheppub} 
                     
\usepackage[T1]{fontenc} 
\usepackage{orcidlink}
\newcommand{\bef}{\begin{figure}}
\newcommand{\eef}{\end{figure}}
\newcommand{\bc}{\begin{center}}
\newcommand{\ec}{\end{center}}

\newcommand{\antinue}{\ensuremath{\overline{\nu}_{e}}}

\newcommand{\comment}[1]{}

\newcommand{\sout}[1]{}

\title{Study of the beyond standard model interaction using Coherent Elastic 
Neutrino-Nucleus Scattering process}

\author[a,b]{S. P. Behera\,\orcidlink{0000-0003-0919-6575}}
\author[a,b]{S. Panda\,\orcidlink{0009-0005-7099-6151}}
\author[a,b]{D. K. Mishra\,\orcidlink{0000-0002-3434-793X}}

\affiliation[a]{Nuclear Physics Division, Bhabha Atomic Research Centre,\\ Mumbai 
- 400085, India}
\affiliation[b]{Homi Bhabha National Institute, Anushakti Nagar, Mumbai - 400094, 
India}

\emailAdd{shiba@barc.gov.in}
\emailAdd{srpanda@barc.gov.in}
\emailAdd{dkmishra@barc.gov.in}

\abstract{
We have conducted an extensive study that highlights the potential of the Indian Coherent
Neutrino-nucleus Scattering Experiment (ICNSE) detector in constraining 
neutrino-quark interactions that go beyond the standard model. By utilizing reactors 
with varied core configurations and power outputs as sources for electron 
antineutrinos, and operating with a target mass of 10 kg over a year, our findings 
reveal that the ICNSE detector is remarkably effective in narrowing down the vast 
majority of the Non-standard Interaction (NSI) parameter space. Moreover, 
incorporating results from two distinct detectors, like sapphire and high-purity 
germanium, markedly enhances sensitivity by 
reducing the degeneracies between some pairs of NSI parameters to a smaller region.
This research highlights its role in enabling future developments and investigations.
}
\begin{document} 
\label{sec:intro}
\maketitle
\flushbottom
The coherent elastic neutrino-nucleus scattering (CE$\nu$NS) process is a low energy 
phenomenon which has been predicted theoretically in the standard model(SM) more than 
five decades ago~\cite{Freedman:1973yd}. At quark level,  it is now generally well 
understood within the framework of the SM of particle physics. However,
there are some uncertainties due to  nuclear-structure such as
neutron form factor~\cite{AristizabalSierra:2019zmy}, spin dependent cross section
 like axial and tensor interaction for lighter nuclei, strangeness, etc. may induce a
modification on theoretical prediction.
Due to low energy of neutrinos($<$100 MeV), the de Broglie 
wavelength of the scattering process is larger than the nuclear radius ($\lambda$ = 
$h/q \geq  R$, where $q$ refers to the exchanged momentum, $R$ is the nuclear 
radius). Therefore, the scattering amplitudes on single nucleons add coherently and 
lead to an enhanced cross section whose value is directly proportional to number of 
neutrons present in the target nuclei. Despite its large cross section, earlier it 
has been difficult to observe the CE$\nu$NS process using most conventional 
high-mass neutrino detectors due to the very small resulting nuclear recoil 
energies. 

Recently, the COHERENT experiment has carried out measurements of the CE$\nu$NS process, 
utilizing  neutrinos produced by an accelerator. They have established this process 
at a confidence level(CL) of 6.7$\sigma$ using a CsI[Na] scintillator 
detector~\cite{COHERENT:2017ipa,COHERENT:2021xmm}. Moreover, they have reported a 
second measurement conducted with a liquid argon (LAr) 
detector\cite{COHERENT:2020iec}. The findings reveal that the measured cross 
section is in agreement with the predictions made by the SM. Additionally, the 
COHERENT group has achieved a groundbreaking measurement of the CE$\nu$NS process, 
attaining a significance of 3.9$\sigma$ with high-purity germanium spectrometers 
that operate at an energy threshold of 1.5 keV~\cite{COHERENT:2024axu}. The 
measurement of the CE$\nu$NS cross section provides a pathway to explore a variety of 
physics phenomena across different research fields including particle physics, 
nuclear physics, astrophysics, and cosmology. Analyzing the CE$\nu$NS process can 
enhance our understanding of essential physics concepts that extend beyond the SM, 
such as non-standard interactions~\cite{Papoulias:2017qdn}, the neutrino magnetic 
moment~\cite{Kosmas:2015sqa}, the weak mixing 
angle~\cite{Canas:2017umu,Papoulias:2017qdn,AtzoriCorona:2023ktl,AtzoriCorona:2025ygn},
 and the distributions of nuclear neutron density~\cite{AtzoriCorona:2023ktl}. 
This process also allows for detailed investigations into the interiors of dense 
objects and stellar evolution~\cite{Biassoni:2011xuo,Brdar:2018zds}. The CE$\nu$NS 
process not only offers to explore the Beyond Standard Model (BSM) physics but also 
can be applied for reactor monitoring and safeguard ~\cite{Bowen:2020unj}. The 
measurement of CE$\nu$NS process 
using reactor antineutrinos offers some advantages compared to neutrino produced due 
to pion-at-rest decay source. Due to low energy( $<$ 10 MeV), the nuclear form 
factor is almost one at these energies. Furthermore, the possible existence of 
sterile neutrinos could be confirmed or disproved by using purely electron 
antineutrinos and observing neutrinos through the flavor-blind (CE$\nu$NS) process. 
The CE$\nu$NS process was demonstrated by the Dresden-II experiment using electron 
antineutrinos from a nuclear reactor~\cite{Colaresi:2022obx}. Furthermore, the 
CONUS~\cite{CONUS:2024lnu} and CONNIE~\cite{CONNIE:2021ggh} 
experiments have established upper limits on CE$\nu$NS that align with the SM 
predictions, and they are currently working towards their initial detection. 
Recently, CONUS+ experiment reports the first observation of elastic
neutrino–nucleus scattering in the fully coherent regime with low-energy neutrinos
produced in nuclear reactors with a statistical significance of 3.7$\sigma$ 
~\cite{Ackermann:2025obx}. Coherent scattering of xenon nuclei in dark matter search 
initiatives, specifically PandaX-4T~\cite{PandaX:2024muv} and 
XENONnT~\cite{XENON:2024ijk}, has led to the observation of solar neutrinos 
($^{8}B$). There are several experiments are going on and some are underway to 
measure this process using neutrinos from different sources.

In India, an experimental setup the Indian Coherent Neutrino-nucleus Scattering 
Experiment(ICNSE) has been proposed to measure the (CE$\nu$NS) cross section using 
electron antineutrinos produced from reactor and address various fundamental physics 
aspects. The current study examines the detection capabilities of the ICNSE detector 
to constrain the possible presence of new type of interaction in beyond the usual V-A
structure within the SM called as Non-Standard Interaction(NSI) of neutrinos. The 
presence of this new interaction creates a diverse phenomenology in both scattering 
and neutrino oscillation experiments~\cite{Proceedings:2019qno}. This is a 
model-independent framework proposes a  new type of contact interactions on the 
neutrino sector that differs from the electroweak theory in SM. In general, the NSI 
can be occurred in neutral current and charged current processes. Both possibilities 
propose the existence of new kind of mediator that carries heavier mass or in the 
same order with the electroweak counterparts. Regarding the CE$\nu$NS process, only 
the neutral current case is concerned. Neutrino within NSI framework affects quark 
constituent of a nucleus, which can be occurred as non-universal flavor conserving or 
flavor violating process.

The examination of sensitivity to the NSI parameters is significant that helps 
understanding neutrino properties, as any affirmative signal may indicate the 
presence of new physics BSM. The potential presence of neutrino NSIs may impact 
neutrino oscillation rates, enhance our understanding of how neutrino masses are 
generated, and open up new paths for exploring dark matter and cosmological 
physics. Conversely, limitations imposed on these parameters could potentially 
eliminate certain models of new physics.

The paper is structured as follows. In the following section, a detailed design 
concept of the proposed ICNSE setup is presented. The CE$\nu$NS process both within 
the SM and in presence of NSI is discussed in Sec.~\ref{sec:cens}. The simulation 
method for estimating expected number of events in the detector is described in 
Sec.~\ref{sec:expect}. The procedure for finding the sensitivity of the proposed 
experiment is described in Sec.~\ref{sec:simul}. The sensitivity of detector to NSI 
parameters for an exposure of one year is elaborated in Sec.~\ref{sec:results}. In 
Sec.~\ref{sec:summary}, observations obtained from this study is summarized  and 
discussed the implication of this work. 
\section{The proposed ICNSE setup}
\label{detSetup}
Experiments based on the CE$\nu$NS process demand detectors that operate at very 
low threshold energy, which are vital for measuring nuclear recoil energies that are 
typically in the few keV range. Semiconductor detectors made from germanium or 
silicon are suitable candidates for meeting such low threshold requirements. 
However, the energy thresholds of these detectors are generally around
 few hundreds eV range (for Ge detector)~\cite{Colaresi:2022obx,CONUS:2024lnu},
 used to effectively observe the CE$\nu$NS process with 
reactor antineutrinos~\cite{Ackermann:2025obx,Colaresi:2022obx}.
Cryogenic detectors present a viable solution, as they can 
achieve both low thresholds and excellent energy resolution. These detectors can 
be fabricated from various materials, including semiconductors like germanium and 
insulators such as calcium tungstate and sapphire (Al$_{2}$O$_{3}$). However, 
sapphire is an excellent candidate for observing CE$\nu$NS because of the lower 
atomic masses of $Al$ and $O$, which enhances sensitivity to lower nuclear recoil 
energies. Sapphire has favorable phononic characteristics, and it has already been 
demonstrated that low-energy thresholds $\mathcal{O}$(100 eV) are achievable with 
these crystals. A baseline recoil energy resolution of 18 eV, which corresponds to a 
recoil energy threshold of 54 eV, has been accomplished using a newly developed 100 g 
sapphire detector~\cite{Verma:2022tkq}. According to a sensitivity analysis conducted 
in Ref.~\cite{Strauss:2017cuu}, the CE$\nu$NS process can be detected at a 5$\sigma$ 
level with cryogenic detectors that have recoil energy thresholds of 20 eV. 
Therefore, in this work, sapphire and germanium detector have been utilized as the target material. The 
conceptual design of the detector setup has been derived from 
Ref.~\cite{Behera:2023llq}. The detector design includes the multilayered structure 
of shielding materials for reducing both natural and reactor related background. 
   The method of signal extraction from the detector is described in 
Ref.~\cite{Behera:2025jaw} and list of references mentioned there.
\section{Theoretical Model on Interaction Process}\label{sec:cens}
The relevant cross sections for CE$\nu$NS process both within and beyond the SM  
scenario under consideration is presented here. In the following section the SM cross 
sections has been discussed followed by the cases of neutrino Non-standard 
Interaction (NSI), and Neutrino Generalized Interaction (NGI).
\subsection{The standard model CE$\nu$NS cross section}
  In the SM, the differential CE$\nu$NS scattering cross section is given by
\begin{equation}\label{eq:xsec}
\frac{d\sigma}{dT}(E_{\nu},T) = \frac{G_{F}^{2}}{\pi} Q_{W}^{V}{^2} 
\times M_N\left(1-\frac{T M_N}{2E_{\nu}^{2}}\right)|F(q)|^2 .
\end{equation}  
In Eq.~\ref{eq:xsec}, $M_N$ is the mass of nucleus. The SM weak charge is expressed as
\begin{equation}
\label{eq:weakcharge}
 \left(Q_{W}^{V}\right)^{2} = \left(g^{V}_{p}Z + g^{V}_{n}N\right)^{2}.  
\end{equation}
In Eq.~\ref{eq:weakcharge}, $N, Z$ are the number of neutrons, and 
number of protons in the nucleus, respectively. In Eq.~\ref{eq:weakcharge}, $g^{V}_{p}$ and  $g^{V}_{n}$ are corresponding coupling 
of proton and neutron, respectively to $Z^{0}$ boson. The coupling terms are 
expressed through weak mixing angle $sin^{2}\theta_{W}$ = 0.2386 by the relation 
given by $g^{V}_{p} = 1/2 -2~sin^{2}\theta_{W}$ and $g^{V}_{n}$ = 
-1/2. Further, $E_{\nu}$ is the incident neutrino energy, $T$ is 
nuclear recoil energy, ($T_{\rm  max}(E_{\nu})=2E_{\nu}^{2}/(M_N + 2E_{\nu})$), $G_{F}$ 
is the Fermi coupling constant, $\theta{_W}$ is the weak mixing angle, and $F(q)$ is 
the nuclear form factor for a momentum transfer of $q$ ($\sim$1.0 at low $q$).
\subsection{The Neutrino Non-standard Interaction}
The CE$\nu$NS process modifies due to the presence of the NSI of neutrinos and 
quarks. At low energies ($\ll M_{Z}$), the vector-type neutrino NSI with u and d 
quarks can be described by the effective Lagrangian
\begin{equation}
\label{eqnsi}
\begin{split}
{\cal L}_{\nu Hadron}^{NSI} & =-2\sqrt{2}G_{F}\sum_{{q=u,d}\atop{\alpha,
\beta=e,\mu,\tau}}\left[\bar\nu_\alpha \gamma^\mu (1-\gamma^5)\nu_\beta\right] \times 
\\
 & \left( \varepsilon_{\alpha\beta}^{qL}\left[\bar q\gamma_\mu (1-\gamma^5) q\right] 
 + \varepsilon_{\alpha\beta}^{qR}\left[\bar q\gamma_\mu (1+\gamma^5) q\right] 
\right),
\end{split}
\end{equation}
where $\varepsilon_{\alpha\beta}^{qP}$ ($q=u,d$, $P=L,R$ and, $\alpha,\beta$ =e, $\mu,\tau$ ) describe the NSI 
parameter, both for the non-universal terms, $\varepsilon_{\alpha\alpha}^{qP}$, as 
well as for the flavor-changing contributions, $\varepsilon_{\alpha\beta}^{qP}$ 
($\alpha\neq\beta$). Parameters $\varepsilon$ measure the strength of interaction in 
analogous to Fermi theory that is given by $\varepsilon_{\alpha\beta} \sim 
\frac{\sqrt{2}}{G_F}\frac{g_{_{X_{\nu}}}g_{_{X_{q}}}}{M_{X}^{2}}$, where $M_{X}$ and $g_{X}$ are 
the mass and coupling strength of a heavy mediator. Due to this new type of interaction 
with heavy mediator, the weak charge is replaced by corresponding NSI charge that is 
expressed as
\begin{equation}\label{eq:nsi}
\begin{split}
\left(Q_{NSI}^{V}\right)^{2}&=
\left[\left(2\varepsilon_{\alpha\alpha}^{uV} +\varepsilon_{\alpha\alpha}^{dV}
+g_{p}^{V}\right)Z 
+\left(\varepsilon_{\alpha\alpha}^{uV}+2\varepsilon_{\alpha\alpha}^{dV} 
+g_{n}^{V}\right)N\right]^{2} \\
&  +\sum_{\alpha\ne \beta}\left[\left(2\varepsilon_{\alpha\beta}^{uV} + 
\varepsilon_{\alpha\beta}^{dV}\right)Z + \left(\varepsilon_{\alpha\beta}^{uV} + 
2\varepsilon_{\alpha\beta}^{dV} \right)N\right]^{2}\,
\end{split}
\end{equation}
where $\varepsilon_{\alpha\beta}^{qV}\equiv \varepsilon_{\alpha\beta}^{qL} + 
\varepsilon_{\alpha\beta}^{qR}$. Also it can be observed from the above 
Eq.~\ref{eq:nsi} that degeneracies arising when two NSI parameters are assumed to 
be non-vanishing at a time. In the above Eq.~\ref{eq:nsi} by setting the 
$\varepsilon$ to zero leads to the SM CE$\nu$NS. 
In this study the neutral current NSI of neutrinos with the vector couplings has 
been considered. Currently, the focus is on investigating the potential for NSI 
arising from electron anti-neutrinos, as the muon NSI parameters are typically more 
constrained by other experiments. Then considering the above mentioned reason, the 
NSI interaction due to electron anti-neutrino, Eq.~\ref{eq:nsi} is modified as
\begin{equation}\label{eq:elec_nsi}
\begin{split}
\left(Q_{NSI}^{V}\right)^{2} &= \left[\left(2\varepsilon_{ee}^{uV} 
+ \varepsilon_{ee}^{dV} 
+g_{p}^{V}\right)Z +\left(\varepsilon_{ee}^{uV} + 2\varepsilon_{ee}^{dV} 
+g_{n}^{V}\right)N \right]^{2}\\
& +\left[\left(2\varepsilon_{e\tau}^{uV} + \varepsilon_{e\tau}^{dV}\right)Z 
+ \left(\varepsilon_{e\tau}^{uV} + 2\varepsilon_{e\tau}^{dV} \right)N\right]^{2} .
\end{split}
\end{equation}
The NSI interaction cross section can be obtained by replacing $Q_{W}^{V}$ by 
$Q_{NSI}^{V}$ in Eq.~\ref{eq:xsec}. The constraint on parameters 
$\varepsilon_{\alpha\beta}$ due to the interaction of electron and muon type  
neutrinos with quarks are summarized in Ref.~\cite{Scholberg:2005qs}. It has been 
observed that both $\varepsilon_{ee}$ and $\varepsilon_{e\tau}$ are poorly 
constrained. Therefore, in this study it is focused on constraining both
$\varepsilon_{ee}$ and $\varepsilon_{e\tau}$.
\subsection{Neutrino generalized interaction}%
In addition to the previously mentioned normal effective vector-type NSI
 framework, more exotic BSM interactions with non-standard Lorentz structures,
  such as scalar, tensor, pseudo-scalar, or axialvector couplings, may be feasible. 
  These interactions are treated as neutrino generalized interactions (NGIs) and 
  considered in the evaluation of the detector’s sensitivity. 
  An effective Lagrangian for heavy mediators can be described as~\cite{DeRomeri:2025csu}  
\begin{equation}
  \label{NGIlagr}
  \mathcal{L}^\mathrm{NGI}_\mathrm{NC}  \supset \frac{G_F}{\sqrt{2}}
  \sum_{X=(S,V,T)} \mathcal{Q}_X  \left(\bar{\nu}_e \Gamma^X P_L \nu_e \right) 
  \left(\bar{\mathcal{A}} \Gamma_X \mathcal{A}\right)\, ,
\end{equation}
where, $\Gamma^X=\{\mathbb{I, \gamma^\mu,\sigma^{\mu\nu} \}}$  
for scalar, vector and tensor type interaction respectively.
 $\sigma^{\mu\nu}= \frac{i}{2}[\gamma^\mu,\gamma^{\nu}]$ and 
 ~$P_L \equiv (1 - \gamma^5) / 2$ is
  the left-handed projection operator. The $\mathcal{A}$ in the expression 
  shows the nucleus.

The NGI can contain the pseudo-scalar and axial-vector parts. 
However, the contributions do not yield observable effects in current
 experimental sensitivities 
 $\left[ \mathcal{O}\left(\frac{T_{\mathcal{N}}}{m_{\mathcal{N}}}\right) \right]$
  and axial-vector interactions do not coherently sum over heavy spin-zero 
  nucleons. This results in negligible contributions to CE$\nu$NS cross-section. 
  Although spin-dependent interactions can arise from nuclei with
nonzero spin such as $\mathrm{{}^{27}Al}$ in a sapphire detector. In this 
study we have omitted both pseudo-scalar and axial-vector interactions.

The parameter \(\mathcal{Q}_X\) in Eq.~\eqref{NGIlagr} represents 
the effective coupling between neutrinos and nuclei for 
scalar (\(X = S\)), vector (\(X = V\)) and tensor (\(X=T\))
 type interactions. These couplings determine the contribution
  of neutrino generalized interactions to the CE\(\nu\)NS cross section, 
  which takes the form:
 \begin{eqnarray}
  \left.\frac{\mathrm d\sigma_{\nu \mathcal{A}}}{\mathrm dT_\mathcal{A}} \right|_\mathrm{CE\nu NS}^\mathrm{NGI} & = &
  \frac{G_F^2 M_N}{\pi}F_{W}^2(\left|\bold q\right|^2)\left[\mathcal{Q}_S^2\frac{M_N T_\mathcal{A}}{8E_\nu^2}+\right. \nonumber \\
 & &\left. \left(\frac{\mathcal{Q}_V}{2} + Q_V^\text{SM}\right)^2\left(1-\frac{M_N T_\mathcal{A}}{2E_\nu^2}-\frac{T_\mathcal{A}}{E_\nu}\right)\right] . \nonumber \\
  \label{eq:xsecNGI}
\end{eqnarray}
In the above Eq.~\ref{eq:xsecNGI}, the scalar and vector charges can be expresses as follows  
\begin{align}
\label{eq:couplingsCa}
    \mathcal{Q}_S  & = C_{\nu S}\sum_{q = u, d} C_{q S}\left( Z  \dfrac{m_p}{m_q} 
f_{T_q}^{(p)}  + N \dfrac{m_n}{m_q} f_{T_q}^{(n)} \right), \\[4pt]
    \mathcal{Q}_V &= \phi \, C_{\nu V} \left[Z (2 C_{u V} + C_{d V}) 
    +  N (C_{u V} + 2 C_{d V}) \right]. 
\end{align}
Here, $m_u=2.2$MeV and $m_{d}=4.7$MeV are the masses of up and down quarks,
 respectively~\cite{ParticleDataGroup:2024cfk}. Also, $m_p$ and $m_n$ 
 represent the masses of proton and neutron, respectively. The value of parameter 
 $\phi$, depends on the choice of model, however in this study 
 it is considered as $\phi=1$. The coefficients $f_{T_q}^{(p)}$ and
  $f_{T_q}^{(n)}$ represent the contributions of the quark mass 
  to the nucleon mass. The values these parameters is taken as follows:
\begin{align*}
f_{T_u}^p &= 0.026\, , & f_{T_d}^p &= 0.038\, , \\
f_{T_u}^n &= 0.018\, , & f_{T_d}^n &= 0.056\, .
\end{align*}
In this analysis, we have not treated the $\nu$-quark coupling 
explicitly, rather taken it in a way of quark-dependent coupling,
 which is defined by $C_X^q = \sqrt{C_{\nu X} \cdot C_{qX}}$. 
 The scalar and vector contributions are presented separately, 
 as well as in a mixed configuration.
At an energy scale below electroweak scale, it is natural to 
consider an effective field theory approach. Tensor interactions,
 having a different chiral structure, do not interfere with the SM
  amplitude and can be encoded in an effective nuclear charge 
  analogous to the weak charge. 
Nevertheless, in low energy processes tensorial interactions are sensitive 
to spin dependent (SD) nuclear response (NR)~\cite{Hoferichter:2020osn} for which 
light-spin-odd nuclei, like $\mathrm{{}^{27}Al}$ in our case, are sensitive to 
SD structure factors. The modified cross-section due to such interactions can be 
written in a model independent way by considering the form factor in the nucleon 
basis rather than in isoscalar/isovector basis~\cite{TEXONO:2025sub,Candela:2024ljb}. The modified cross-section can be 
written as: 
\begin{align}
\left.\frac{d\sigma }{dT_{\mathcal{A}}} \right|_{T}^{\text{NSI}}
&= \frac{G_{F}^{2} M_{\mathcal{N}}}{\pi } \Bigg(\frac{8 \pi}{2J+1}\Bigg) \notag 
\times \bigg\{ 
\Big[\varepsilon_{\alpha \beta}^{uT} 
    \Big(\delta_u^p \sqrt{S_p^{\mathcal{T}}(q^2)}
        + \delta_u^n \sqrt{S_n^{\mathcal{T}}(q^2)}\Big)  \notag \\[-3pt]
&\qquad
+ \varepsilon_{\alpha \beta}^{dT} 
    \Big(\delta_d^p \sqrt{S_p^{\mathcal{T}}(q^2)}
        + \delta_d^n \sqrt{S_n^{\mathcal{T}}(q^2)}\Big)
\Big]^2 \,\Big(1 - \frac{M_{\mathcal{N}} T_{\mathcal{A}}}{2 E_{\nu}^2}               
                         - \frac{T_{\mathcal{A}}}{E_{\nu}}\Big) \notag \\[6pt]
&\quad+ 
\Big[
  \varepsilon_{\alpha \beta}^{uT} 
    \Big(\delta_u^p \sqrt{S_p^{\mathcal{L}}(q^2)} 
        + \delta_u^n \sqrt{S_n^{\mathcal{L}}(q^2)}\Big) 
+ \varepsilon_{\alpha \beta}^{dT} 
    \Big(\delta_d^p \sqrt{S_p^{\mathcal{L}}(q^2)} \notag \\[-3pt]
    &\qquad
        + \delta_d^n \sqrt{S_n^{\mathcal{L}}(q^2)}\Big)
\Big]^2 
\Bigg( 2 - \frac{2T_{\mathcal{A}}}{E_{\nu}} \Bigg)
\bigg\},
\label{eq:dSdT_T_eps}
\end{align}
Where, $J$ corresponds to the total angular momentum,
 $S_{n,p}^{\mathcal{T,L}}(q^2)$ are the nuclear response for transverse and 
 longitudinal modes written in proton/neutron basis and  $\delta_q^{(N)} $ 
represent the tensor-charge of the quark in nucleon (proton or neutron). The values 
are $\delta_u^p = \delta_d^n = 0.784$ and  $\delta_d^p = \delta_u^n = -0.204$. 
In all the cases NR can be  written in terms of SD structure factors as a combination 
of $\mathcal{F}_+^{\Sigma'_L}(q^{2})$ and $\mathcal{F}_+^{\Sigma''_L}(q^{2})$, 
which can be written in the general form $\mathcal{F}_{\pm}^{\mathcal{T,L}}(q^{2}) 
=  \mathcal{F}_{p}^{\mathcal{T,L}}(q^{2}) \pm 
\mathcal{F}_{n}^{\mathcal{T,L}}(q^{2})$ ~\cite{Hoferichter:2020osn}. In 
 Eq.~\ref{eq:dSdT_T_eps}, for the simplicity we have neglected 
 $\mathcal{O}(T_{\mathcal{A}}^2)$ and  $\mathcal{O}(\delta a^P(q^2))$ terms from  
our calculation. Comprehensive calculations, including all $L$ modes, have been 
reported previously for the $\mathrm{{}^{27}Al}$ nucleus~\cite{Hoferichter:2022mna}.
In this study the focus is to constraint both
$\varepsilon_{ee}^{uT}$ and $\varepsilon_{ee}^{dT}$. 
The procedure for estimating the expected number of events in the detector as well 
as the possible sensitivity of the detector to NSIs parameters that the proposed 
experiments can reach are described in the next section.
\begin{table*}[t]
\small 
 \begin{center}
\caption{\label{tab:reactortype}{Various types of reactors used as $\antinue$ sources }}
\begin{tabular}{ cccc}
    Reactors name & Thermal power(MW$_{th}$) & Fuel type & Core sizes,
     R: radius, H: Height\\
\hline
   Apsara-U & 2.0 & U$_{3}$Si$_2$-Al (17$\%$ enriched $^{235}$U)& R = 0.32 m, H = 0.64 m \\
    Dhruva & 100.0 & Natural uranium (0.7$\%$  $^{235}$U) & R = 1.5 m,~ H =  3.03 m \\
 PFBR & 1250.0 & MOX(PuO$_{2}$-UO$_{2}$) & R = 0.95 m, H =  1.0 m \\ 
 VVER & 3000.0 & UO$_{2}$ (3.92 $\%$ enriched $^{235}$U)&  ~~R = 1.58 m,  H = 3.53 m \\
   \hline
\end{tabular}
\end{center}
\end{table*}
\section{Simulation Method for Estimating the Expected event rate and Extraction of 
the Detector Sensitivity}
\label{sec:expect}
Event rate has been estimated considering detection energy threshold of 100 and 150 eV for
sapphire and germanium detector, respectively, 
80$\%$ of the detection efficiency independent of nuclear recoil, 90$\%$ fiducial 
volume of the detector, 70$\%$ reactor duty cycle and, for an exposure of  one year. 
Studies have been conducted using a target mass of 10 kg to extract the detector 
sensitivity for various parameters related to neutrinos NSI process. The number of 
events expected in the sapphire detector due to various reactor power as well as 
reactor core to detector distance has been estimated using the CE$\nu$NS process and 
mentioned in Ref.~\cite{Behera:2025jaw}.  The number of
events expected in the detector has been evaluated by knowing the energy dependent
 flux, cross-section, the detector efficiency, fiducial volume of the detector and the reactor duty
cycle. The production point of electron antineutrinos inside an extended reactor core  are 
randomly generated while it is assumed to be a  point detector due to its small size.
\begin{figure*}[t]
\centering
\includegraphics[width=0.4\textwidth]{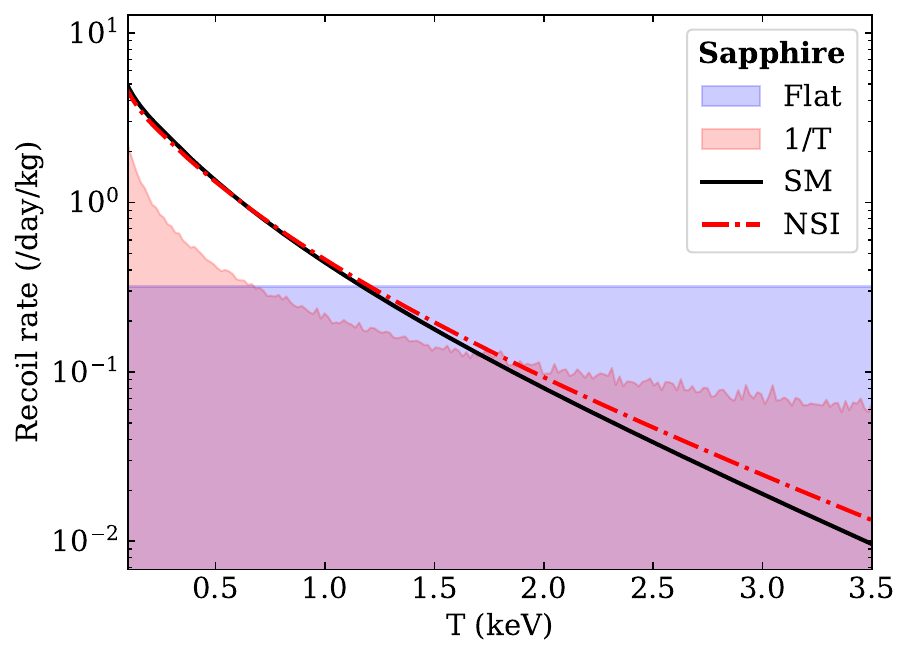}
\includegraphics[width=0.4\textwidth]{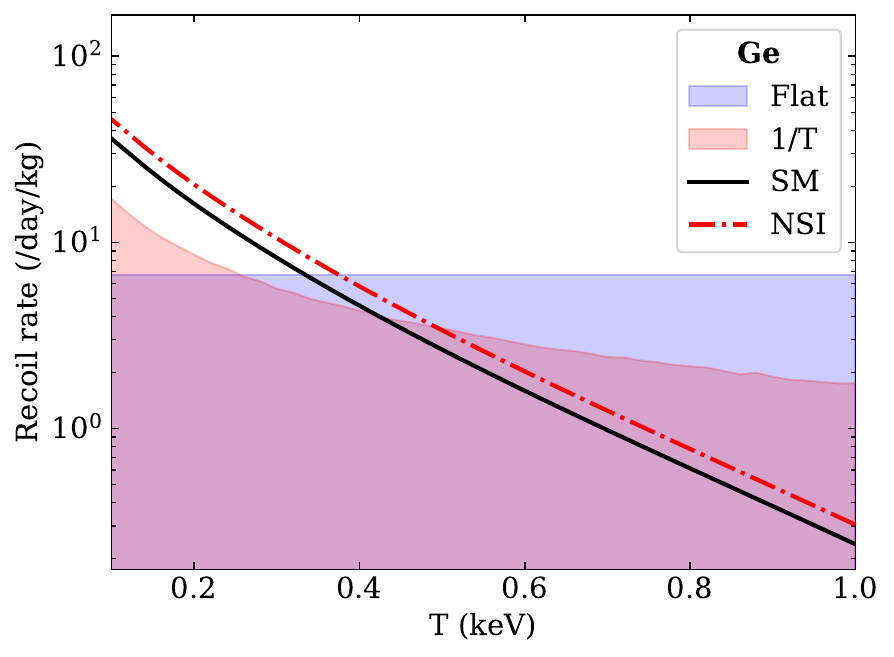}
\caption{ \label{fig:eventrate}Recoil event rate and different backgrounds of 
various detectors such as $Al_{2}O_{3}$(left panel) and Ge(right panel) place at 4m 
distance from Apsara-U reactor. The rate has been estimated for an exposure of one 
day with target mass of 1 kg.} 
\end{figure*}
 The procedure for finding the event in detector is mentioned below
\begin{equation}\label{evt}
\begin{split}
 N^{\rm events} &=\epsilon t f_{0} \frac{M_{\text{detector}}}
 {A}\int_{E_{\nu \rm min}}^{E_{\nu \rm max}}f(E_{\nu})dE_{\nu}\\
&\int_{T_{\rm min}}^{T_{\rm max}(E_{\nu})}\left(\frac{d\sigma}
 {dT}\right) dT ,
\end{split}
\end{equation}
\noindent where $M_{\text{detector}}$ is the mass of the detector, $t$ is the time 
duration of data taking, $f_{0}$ is the total neutrino flux, $f(E_{\nu})$ is the 
neutrino spectrum and, $\epsilon$ is the efficiency of detector. The production 
mechanism of \antinue produced from a reactor is discussed in 
Ref.~\cite{Behera:2023llq}. The minimum recoil energy of the nuclei in a specific 
experimental setup is determined by the detector threshold. Since there are fewer 
neutrinos with higher energy, we have chosen to analyze neutrinos 
with a maximum energy of approximately 6.0 MeV. 
Initially, it is planned to use Apsara-U research reactor facility at Bhabha Atomic 
Research Centre (BARC), India~\cite{SINGH2013141} as a source for \antinue s. The 
Apsara-U reactor's primary benefit is its small and movable core. One advantage of 
using a movable core is that position-specific observations may be used to cancel 
systematic errors related to the reactor and detector. In the later stage the same 
detector setup can be used in other reactor facilities, including Dhruva, 
BARC~\cite{AGARWAL2006747}, proto-type fast breeder reactor 
(PFBR)~\cite{CHETAL2006852}, IGCAR, Kalpakkam, and VVER, Kudankulam in 
India~\cite{AGRAWAL2006812}, might eventually have the same detector configuration. 
Table~\ref{tab:reactortype} provides information on reactor thermal power, 
core sizes, and fuel compositions.
The comparison of recoil event rate due to various types of detector placed at 4 m 
distance from Apsara-U reactor is shown in Fig.~\ref{fig:eventrate}. 
Event rates are 
obtained considering $\varepsilon_{ee}^{uV}$ = $\varepsilon_{ee}^{dV}$ 
=$\varepsilon_{\tau e}^{uV}$=$\varepsilon_{\tau e}^{dV}$=0.1. In the case of 
detector made up of Ge the event rate due to NSI is more compared to event rate due 
to SM CE$\nu$NS process. The increase in event rate is due to the NSI charge 
$Q_{NSI}^{V}$. It is found that the presence of non-zero NSI can increase or reduce 
the event rate as shown in left panel of Fig.~\ref{fig:eventrate}. This observation
shows that the NSI phenomenon is a good candidate to extract new physics from the BSM 
phenomenon. The measurement 
of  low energy of recoil nuclei from the CE$\nu$NS process faces a significant 
challenge due to the background associated with the reactor, including neutron 
and gamma radiation, along with the cosmogenic background such as muon-induced neutrons, 
cannot be entirely removed even with shielding, will interfere with the actual signal.
The detector's sensitivity is affected by both the types 
of backgrounds and their energy-dependent configuration. At low recoil energy, two 
distinct background  shapes are observed: the $1/T$ shape and flat-shaped 
backgrounds, as discussed in Ref.~\cite{Bowen:2020unj,Strauss:2017cuu}. By considering a signal-to-background ratio
one, individual background rate are also shown for comparison.
\subsection{Extraction of the Detector Sensitivity }
\label{sec:simul}              %
This study aims to evaluate the capabilities of the ICNSE detector in measuring
 NSI parameters through the use of \antinue s produced from different reactor 
facilities, as outlined in Table~\ref{tab:reactortype}. The parameterization 
of the \antinue s flux due to various fuel compositions of different reactors  is 
detailed in Ref.~\cite{Behera:2023llq}. These \antinue s are resulting from the 
beta decay of fission fragments, specifically focusing on energy spectra above 2.0 
MeV and the corresponding parameterization has been considered from 
Ref.~\cite{Huber:2011wv,Mueller:2011nm}. Similarly, the low energy \antinue s 
produced due to fission has been considered and it is mentioned in 
Ref.~\cite{Behera:2023llq}.
Further, \antinue s produced due to thermal  neutron capture by the $^{238}U$ exhibit 
energies below 2 MeV, and corresponding numerical data for this segment of the 
spectrum has been sourced from Ref.~\cite{TEXONO:2006xds}. To assess the detector's 
sensitivity to NSI parameters, the analysis presumes that an experiment aimed at 
detecting CE$\nu$NS events will align with the SM predictions. Any 
discrepancies would suggest the existence of a new physics. A statistical analysis (rate-only) is 
conducted to compare the predicted event distributions from simulations with the 
expected distributions, thereby quantifying the detector's sensitivity for a 
specific exposure. The sensitivity to various parameters is subsequently determined 
by calculating the $\chi^2$ value.
The definition of $\chi^{2}$ is taken from Ref.~\cite{Lindner:2016wff} and given as
\begin{equation}
\chi^{2}=\frac{[(1+\xi)N^{th}(\xi)-N^{ex}]^{2}}{\sigma_{{\rm stat}}^{2}
+\sigma_{{\rm sys}}^{2}} +\frac{\xi^{2}}{\sigma_{\xi}^{2}},\label{eq:chi1}
\end{equation}
where $\xi$ denotes the pull parameter with uncertainty $\sigma_{\xi}$. In 
Eq.~\ref{eq:chi1}, $N^{ex}$, $N^{th}$ are representing the number of events obtained 
from the simulations with the  deviation from SM CE$\nu$NS cross section (considered 
as measured) and with consideration of the SM CE$\nu$NS cross section ( considered as 
theoretically predicted) events, respectively. The procedure for estimating 
theoretically predicted events $N^{th}$  with consideration of reactor as well as 
detector related parameters is mentioned earlier. It can be noted here that a single
energy bin is considered for recoil energy range of interest. 
In both types of simulated events, detector response such as resolution and 
efficiency are incorporated. The procedure for detector response incorporation is 
mentioned in Ref.~\cite{Behera:2023llq}. It can be further noted that due to  
rate-only analysis, the effect of detector resolution is negligible.
The statistical uncertainty $\sigma_{{\rm 
stat}}$ and the systematic uncertainty $\sigma_{{\rm sys}}$ of the event
number are given by
\begin{equation}
\sigma_{{\rm stat}}=\sqrt{N^{th}+N_{{\rm bkg}}}\,,\,\,\sigma_{{\rm sys}}
=\sigma_{\alpha}(N^{th}+N_{{\rm bkg}})\
\end{equation}
Here $N_{{\rm bkg}}$ is the number of background events. We assume that 
$\sigma_{{\rm sys}}$ is proportional to the event number with a coefficient 
$\sigma_{\alpha}$. The $\chi^2$ is minimized with respect to pull variables $\xi$ and 
it is estimated by considering different sources of systematic uncertainties. It 
includes  normalization uncertainty(5$\%$) which arises due to reactor total neutrino flux, 
number of target atoms, and detector efficiency, uncertainty due to nonlinear energy 
response of the detector(1.0$\%$), and uncertainty in the energy calibration(1.0$\%$). So an overall 
$\sigma_{\xi}$ = 5.5$\%$ systematic uncertainty has been considered~\cite{Behera:2025jaw}. 
The systematic  uncertainty due to 
backgrounds is considered as $\sigma_{f}$ = 5$\%$.  At different signal-to-background 
ratio has been taken into  consideration to determine the sensitivity of the 
detector in the presence of background.  In the following sections, the sensitivity 
of sapphire detectors to various parameters is discussed.
\section{Results and Discussions}\label{sec:results}
In this section the potential sensitivity of the detector to NSI and NGI parameters 
coming from the coherent neutrino-nuclei scattering process is presented.  It has 
been assumed a target mass of 10 kg and an exposure of 1 year, considering both 
absence and presence of background and with electron antineutrinos as a
probe. In the case of NSI, three different cases are 
studied such as non universal flavor conserving, flavor changing and a combination of
both while limiting the corresponding parameters.  The signature of NSI can
 be inferred due to a deviation from the expected SM 
cross section. Similarly, the sensitivity of the detector due to different NGI 
parameters such as scalar, vector and tensor are discussed below.
\begin{figure*}[th]
\centering
\includegraphics[width=0.34\textwidth]{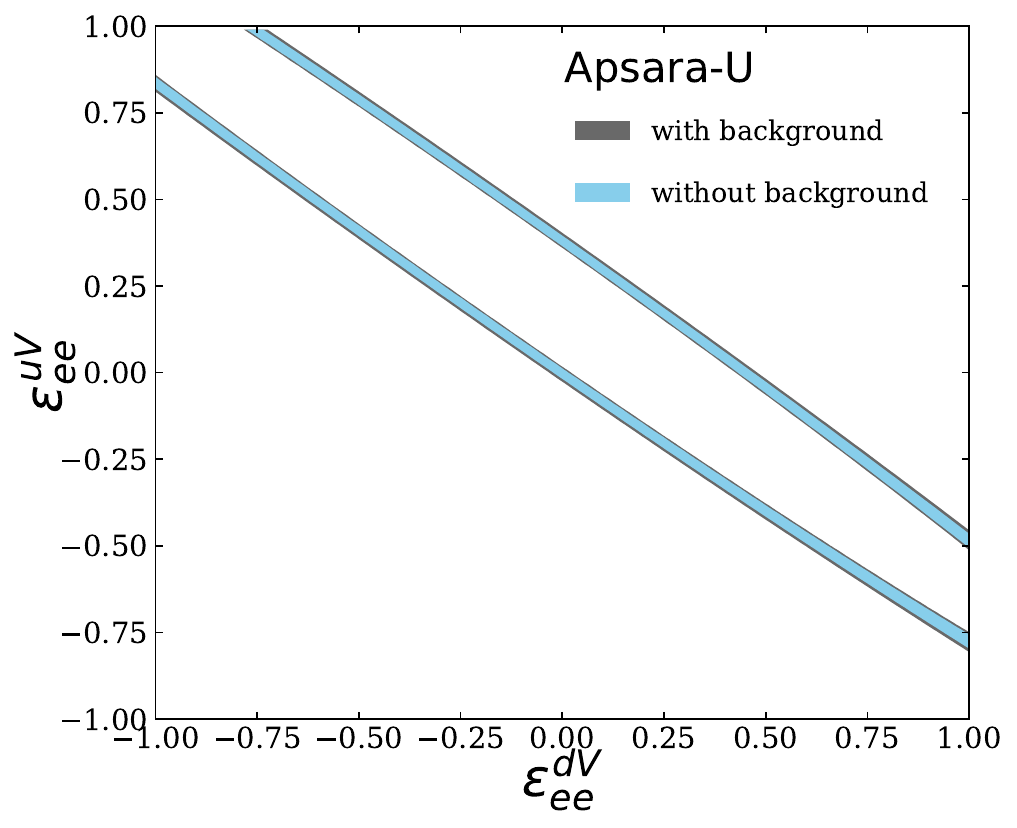}
\includegraphics[width=0.34\textwidth]{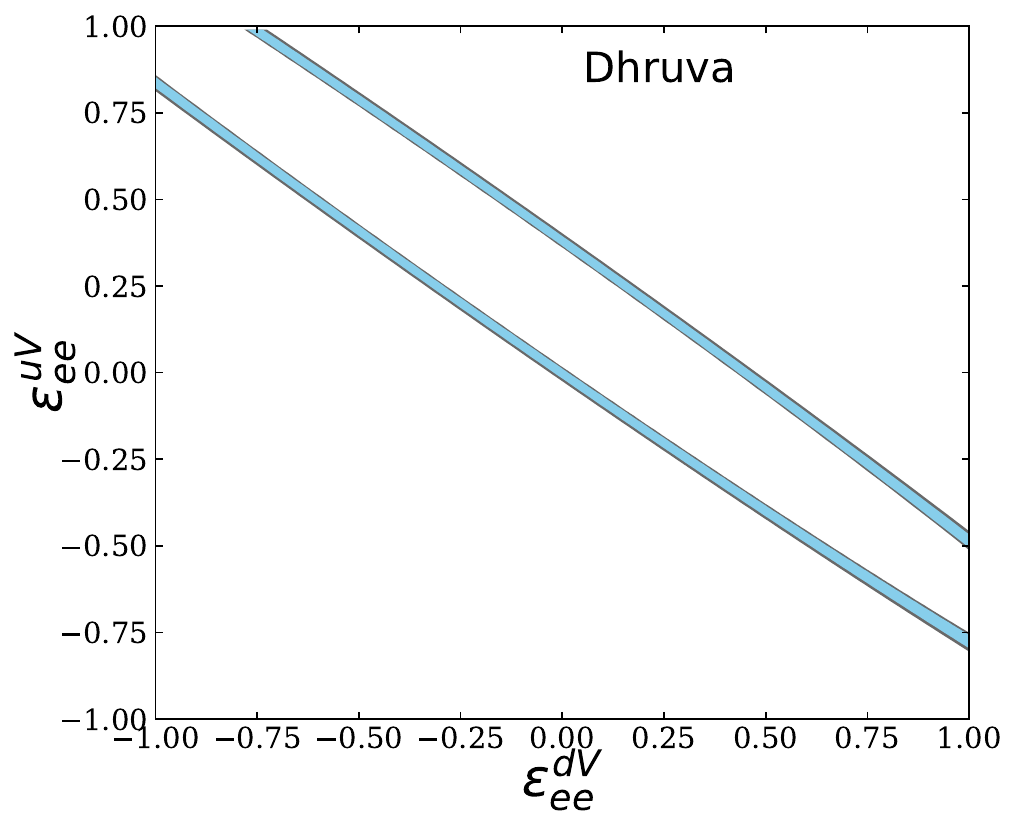}
\includegraphics[width=0.34\textwidth]{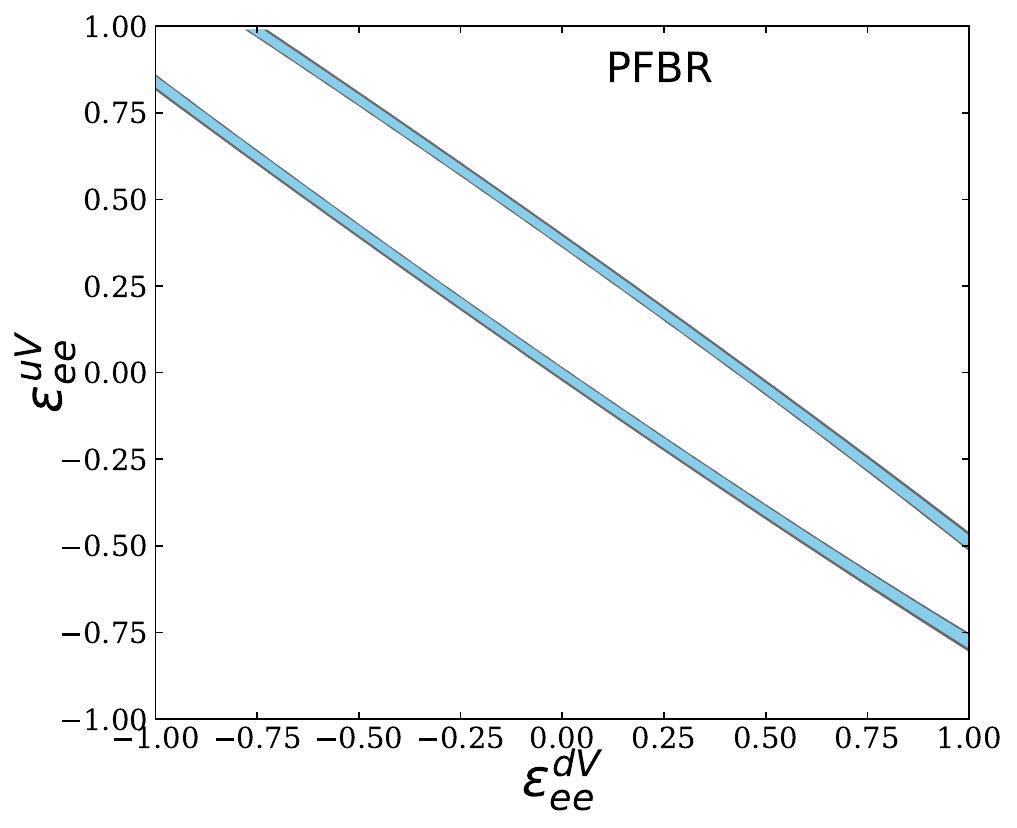}
\includegraphics[width=0.34\textwidth]{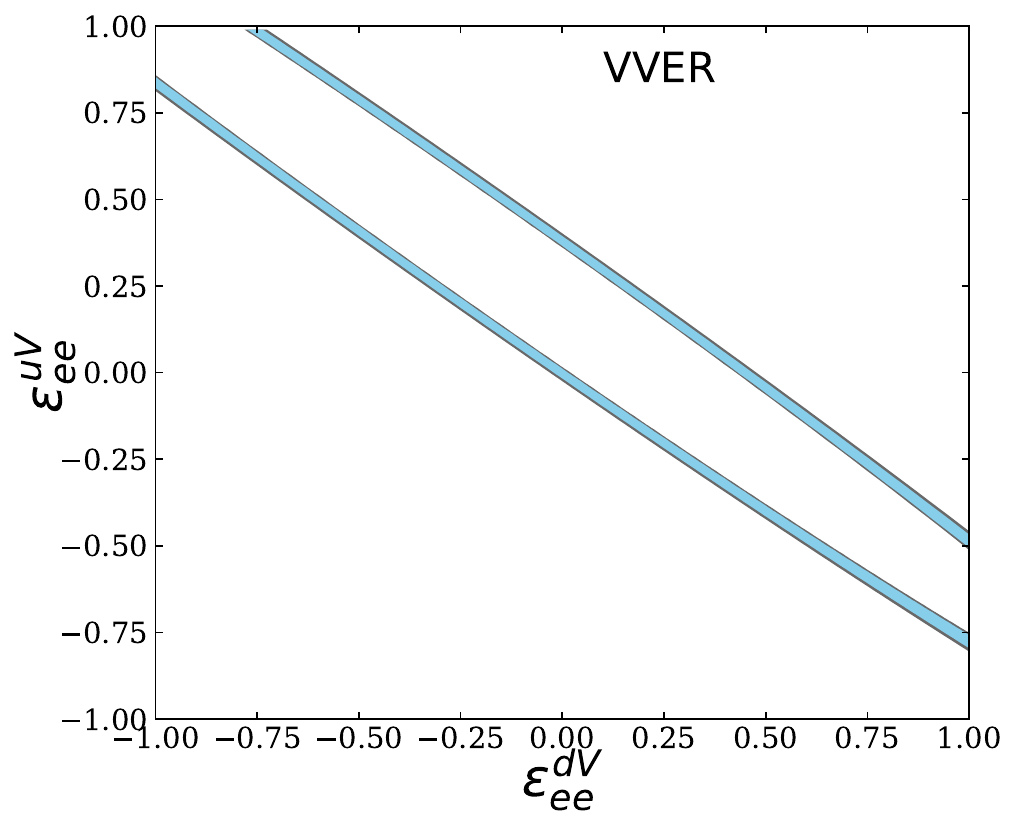}
\caption{ \label{fig:eeud}The allowed regions of non-universal
 NSI parameters $\varepsilon_{ee}^{uV}$-$\varepsilon_{ee}^{dV}$
plane at 90$\%$ CL considering the Apsara-U, Dhruva, PFBR and, VVER reactors as 
antineutrinos sources.} 
\end{figure*}
\begin{figure}
\centering
\includegraphics[width=0.4\textwidth]{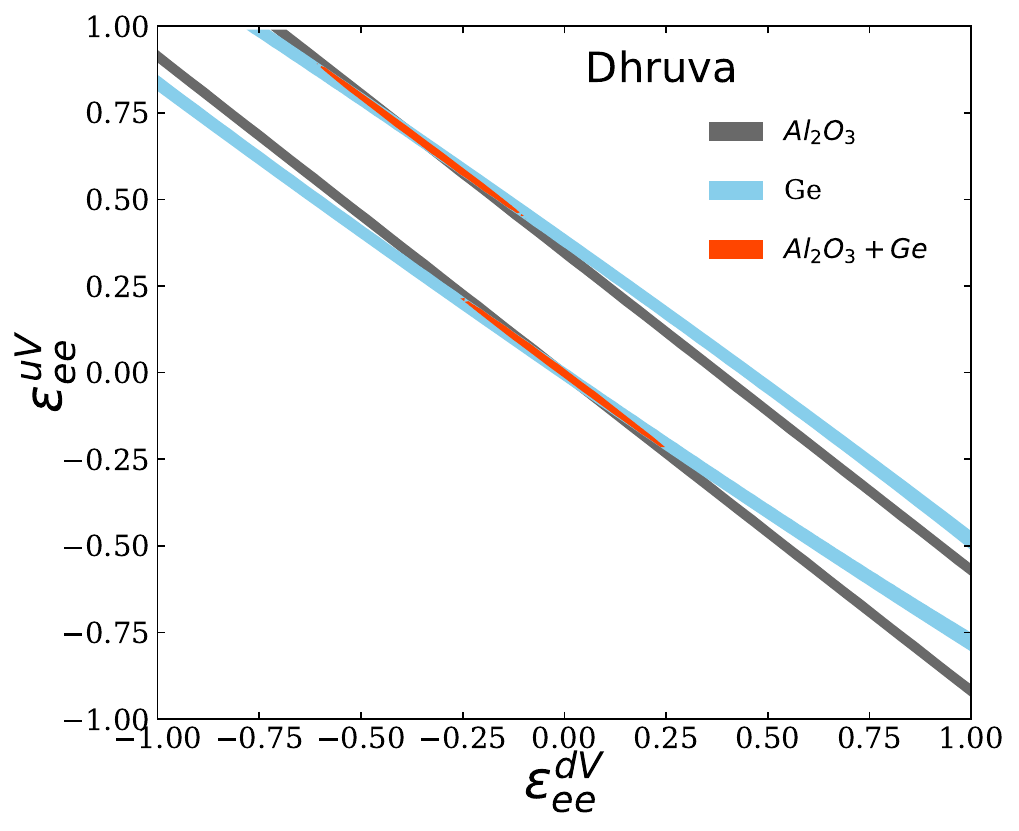}
\caption{ \label{fig:ue_AlGe} An expected  allowed regions 
of non-universal NSI parameters $\varepsilon_{ee}^{uV}$-$\varepsilon_{ee}^{dV}$
plane at 90$\%$ CL considering detector made up  with $Al_{2}O_{3}$ and Ge. 
Red shaded region shows the combined results. Results are extracted 
considering a signal-to-background ratio of 1.} 
\end{figure}
\begin{figure}
\centering
\includegraphics[width=0.4\textwidth,height=0.36\textwidth]{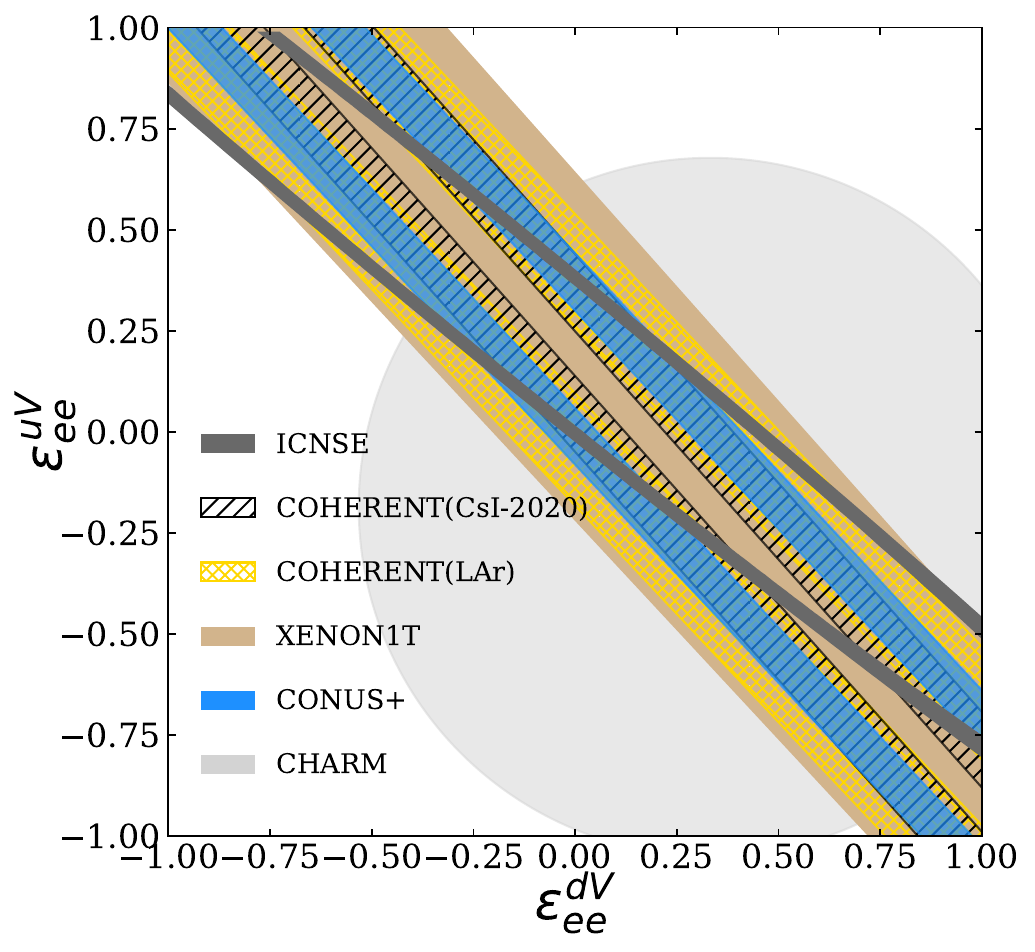}
\caption{ \label{fig:ue_global} In the $\varepsilon_{ee}^{uV}$-$\varepsilon_{ee}^{dV}$
plane at 90$\%$ CL comparisons of sensitivity of the ICNSE detector placed at 10 m 
from Dhruva reactor with others experimental observations.} 
\end{figure}
\subsection{Sensitivity of the detector to non-universal NSI parameters}
The potential reach of the detector has been extracted considering of only 
non-universal parameters. The constraint on NSI parameters obtained considering the 
interaction of neutrinos with both u- and d-type quarks. Figure~\ref{fig:eeud} shows 
expected allowed region in $\varepsilon^{dV}_{ee}-\varepsilon^{uV}_{ee} $ plane
  at 90$\%$ CL for a sapphire detector considering the antineutrino produced   
from various reactors as mentioned earlier. The sensitivity of the detector has 
been obtained considering without and with presence of background and for the 
latter case a signal-to-background ratio is considered as 1.0. It is found that 
both u- and d-types NSI parameters are strongly correlated. The behavior of this 
results can be explained considering the expressions of the weak nuclear charge 
due to the SM CE$\nu$NS process and the weak charge due to the NSI that leads to 
different event rates. Total number of expected events can be estimated by 
considering $\varepsilon^{uV}_{ee}$ and $\varepsilon^{dV} _{ee}$ as mentioned below
\begin{equation}
N^{th} = \left [ Zg_V^{p} + Ng_V^n + \left ( A + Z \right )\varepsilon^{uV}_{ee} 
+ \left ( A + N \right )\varepsilon^{dV} _{ee}\right ]\,
\label{eq:slope}
\end{equation}
where the slope of the curve on the plane ($\varepsilon_{ee}^{uV}$,
$\varepsilon_{ee}^{dV}$) is described by the ratio $m=(A + N)/(A + Z)$.
Due to the above relation mentioned in Eq.~\ref{eq:slope}, allowed regions of 
Fig.~\ref{fig:eeud} appear as a linear band in the 
$\varepsilon^{uV}_{ee}-\varepsilon^{dV}_{ee}$ plane. It has been also found that 
detector sensitivity is almost independent of reactor core configuration and the 
power of the reactor. Further, a more stringent limit on coupling parameters can be 
placed by performing the experiment with a combination of target 
materials~\cite{Barranco:2005yy}. It is also be seen from above that more 
$(A+N)/(A+Z)$ ratio differs between the two targets, the better stringent limits can 
be obtained. In this study, high purity germanium detector also has been considered. 
It has been assumed that the procedure for signal extraction from the detector is 
same as the sapphire detector. Figure~\ref{fig:ue_AlGe} shows the excluded region considering targets of 
$Al_{2}O_{3}$ and $Ge$, each one has mass of 5 kg and with consideration of a 
signal-to-background ratio of 1.0. It can be noted here that, total mass as well as 
exposure have been kept as same while considering a combination of target materials. 
Gray and light-blue shaded portion show the sensitivity  of individual $Al_{2}O_{3}$ 
and $Ge$ detector, respectively for an exposure of one year. Red shaded 
region shows the combined results that improves the sensitivity of detector and 
minimizing the degeneracy in the parameter space.  A similar improvements
 in sensitivity has been found while placing the detector at other reactor 
facilities. Further, it can be noted here that Ge detector sensitivity 
reduces due to quenching factor and the uncertainty associated with it. 
Unless otherwise mentioned, in the present study we have not considered 
both effects while extracting the Ge detector sensitivity.
 Figure~\ref{fig:ue_global} shows the comparison between 
 the expected sensitivity of the ICNSE sapphire detector placed at 10 m from Dhruva 
reactor core and other experimental observations such as results from 
CsI~\cite{COHERENT:2021xmm}, LAr~\cite{COHERENT:2020iec} and 
HPGe~\cite{CONUS:2024lnu} detectors of the 
COHERENT, XENON1T~\cite{XENON:2020gfr}, CHARM~\cite{DORENBOSCH1986303}, and 
CONUS~\cite{CONUS:2021dwh,CONUS:2022qbb} groups. It is found that the sensitivity 
of the ICNSE detector is comparable with other and can put a stringent limit in the 
parameter space.
\begin{figure*}[t]
\centering
\includegraphics[width=0.34\textwidth]{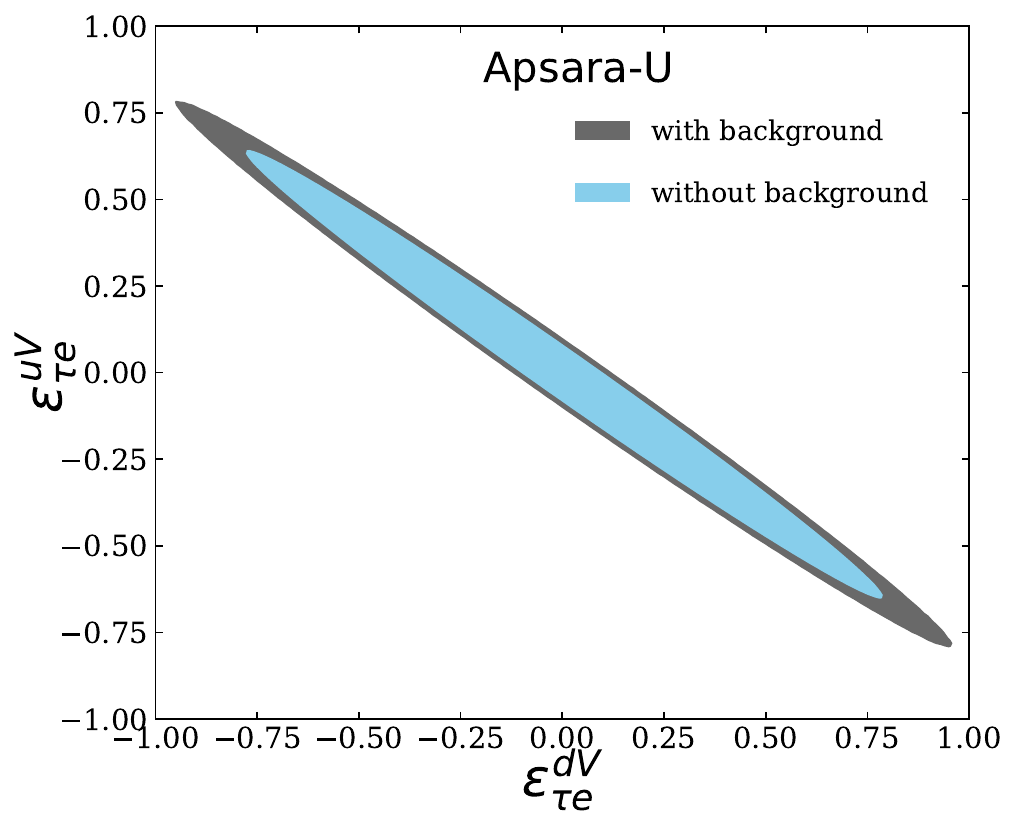}
\includegraphics[width=0.34\textwidth]{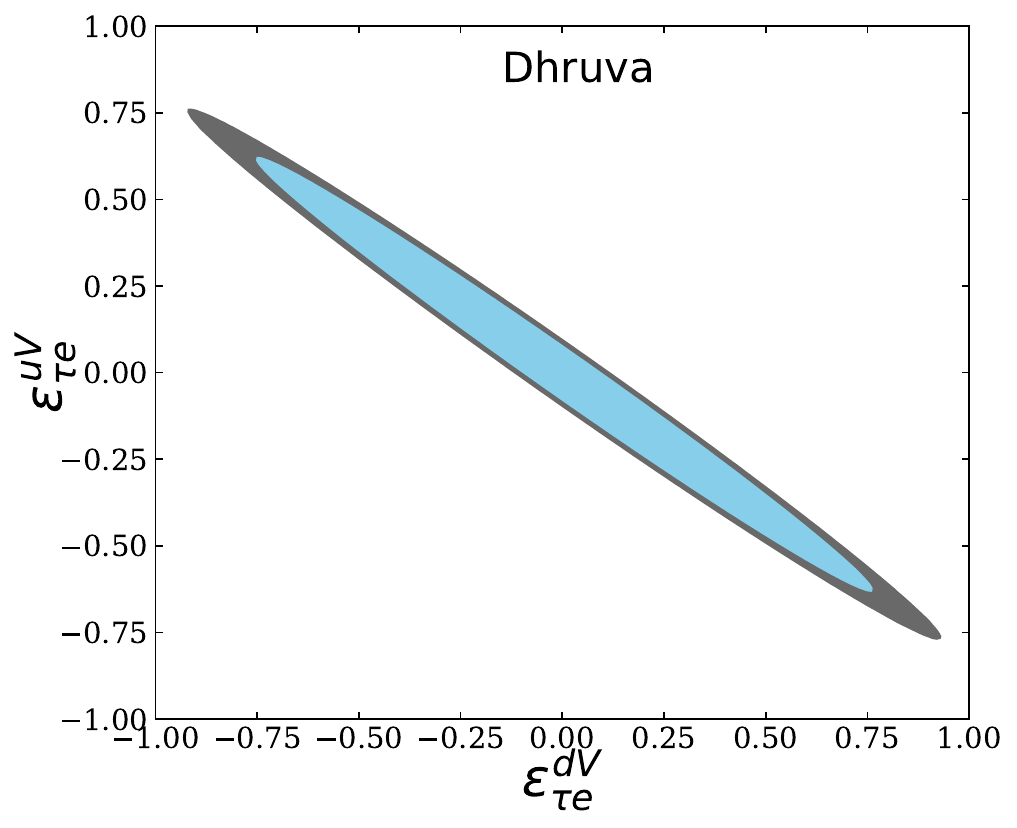}
\includegraphics[width=0.34\textwidth]{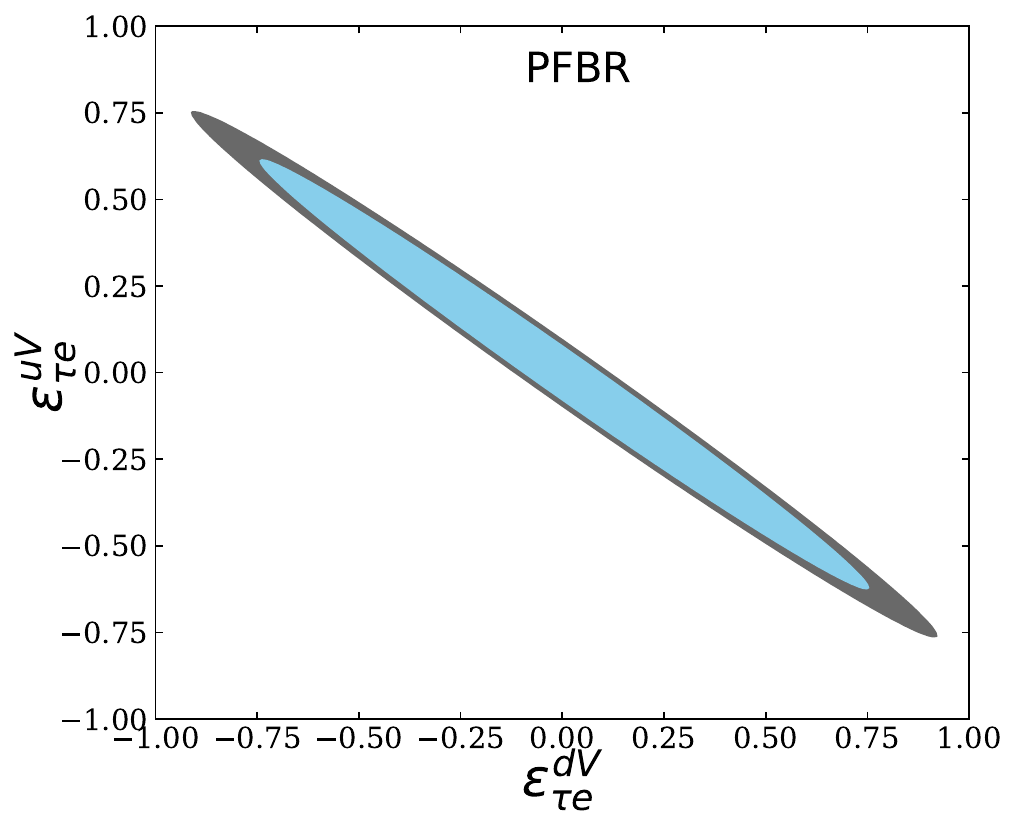}
\includegraphics[width=0.34\textwidth]{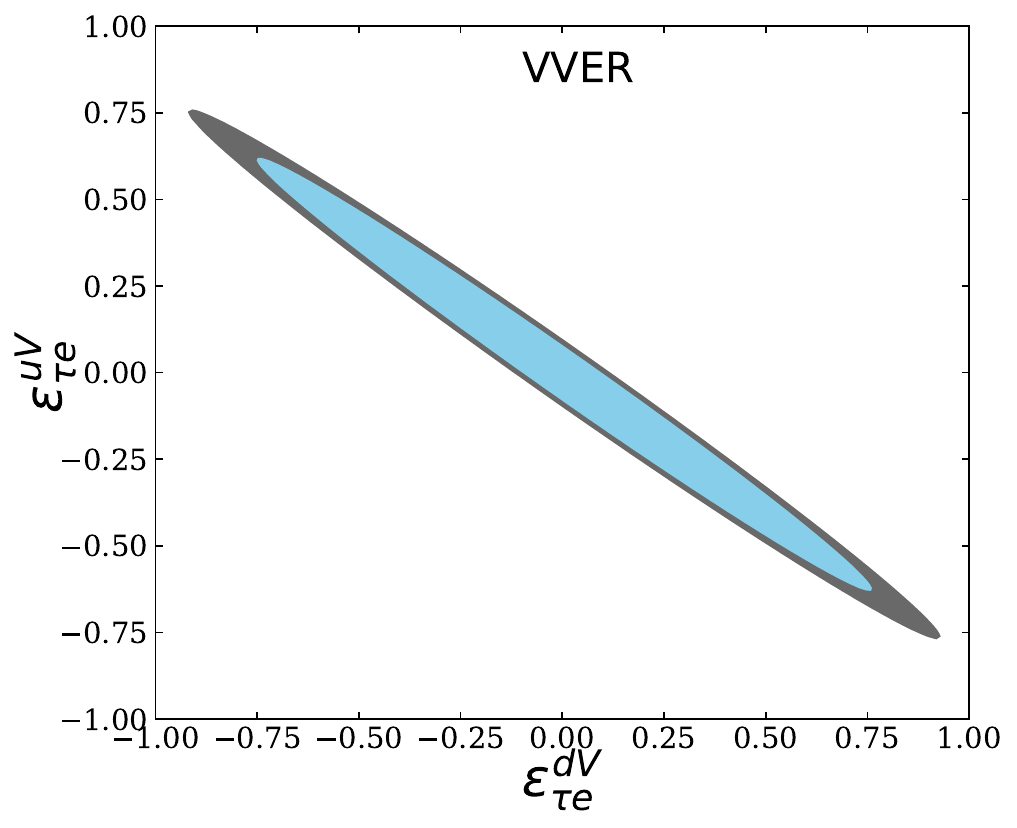}
\caption{ \label{fig:taue} The allowed regions of flavor changing NSI parameters 
$\varepsilon_{\tau e}^{uV}$-$\varepsilon_{\tau e}^{dV}$ plane at 90$\%$ CL 
considering Apsara-U, Dhruva, PFBR and, VVER reactors as antineutrinos sources.} 
\end{figure*}
\begin{figure}[h]
\centering
\includegraphics[width=0.4\textwidth]{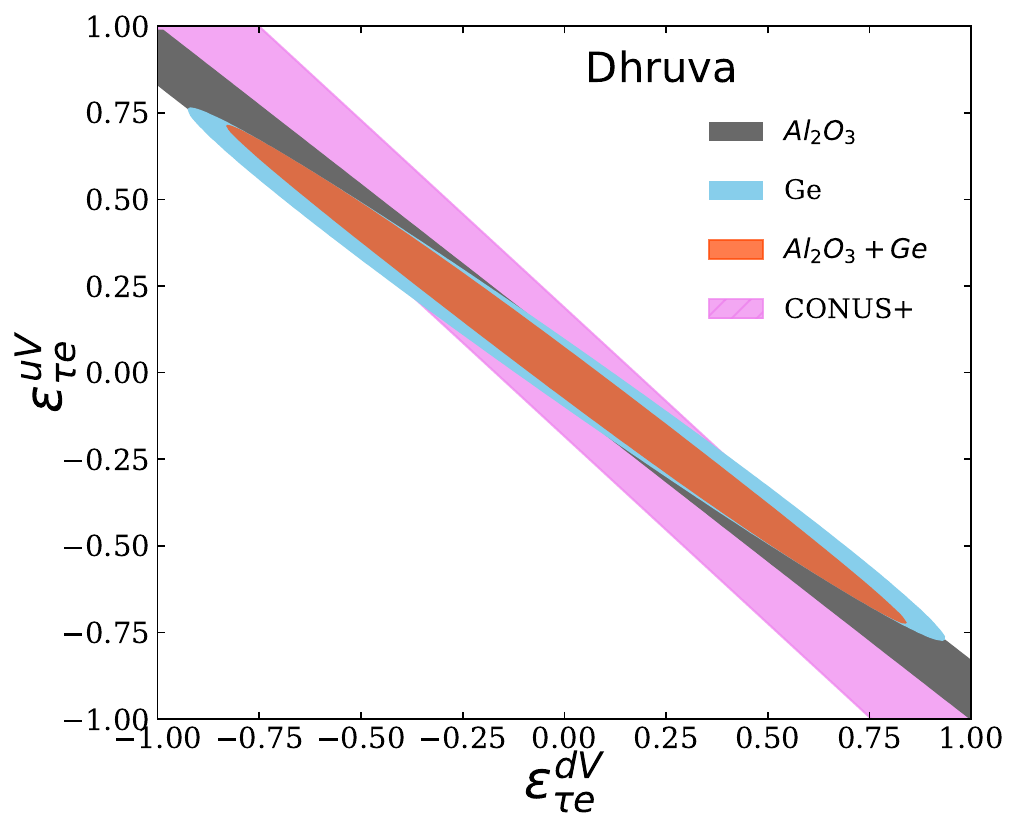}
\caption{ \label{fig:tau_AlGe} An expected  allowed regions of non-universal NSI 
parameters in $\varepsilon_{\tau e}^{dV}$-$\varepsilon_{\tau e}^{uV}$ plane at 
90$\%$ CL considering detector made up  with $Al_{2}O_{3}$ and Ge. Orange shaded
region shows the combined results. Results are extracted considering a 
signal-to-background ratio of 1.0.} 
\end{figure}
\subsection{Sensitivity of the detector to flavor changing NSI parameters}
The sensitivity of the detector has been evaluated with a focus on flavor-changing
 neutrino NSI parameters, specifically excluding non-universal terms. 
 In particular, parameters like $\varepsilon_{\tau e}^{uV}$ and $\varepsilon_{\tau e}^{dV}$
 are considered as in reactor experiment based on CE$\nu$NS both $\varepsilon_{e\tau }$
  and $\varepsilon_{e\mu}$ can be  constrained with comparable strength.
  Figure~\ref{fig:taue} illustrates the 
 expected performance of a sapphire detector in the 
 $\varepsilon^{dV}_{e\tau}-\varepsilon^{uV}_{e\tau}$ 
 parameter space, expressed at a 90$\%$ CL, both when background 
 present and when it is absent. It is notable that the detector's sensitivity is 
\begin{figure*}[ht]
\centering
\includegraphics[width=0.34\textwidth]{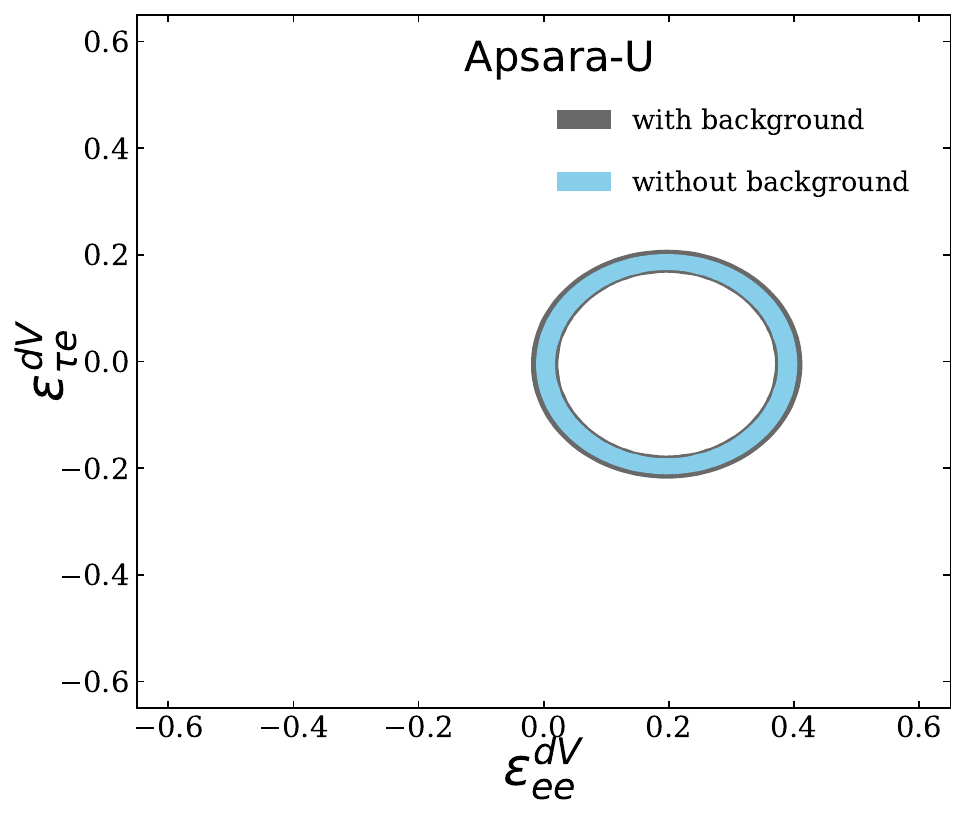}
\includegraphics[width=0.34\textwidth]{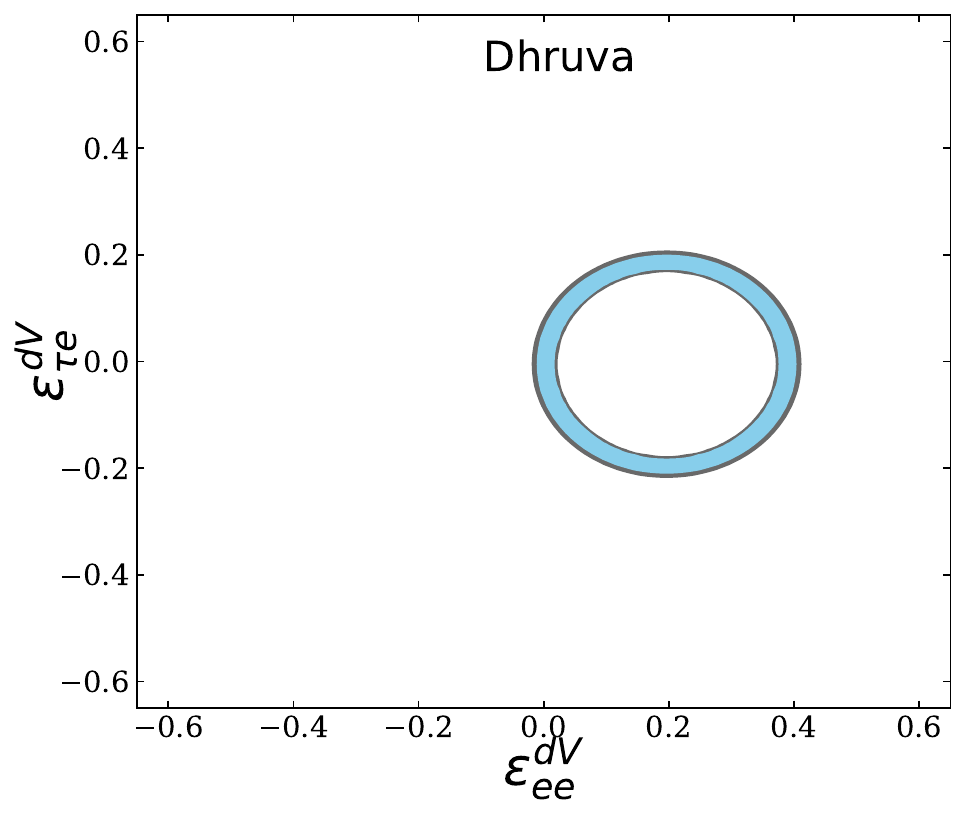}
\includegraphics[width=0.34\textwidth]{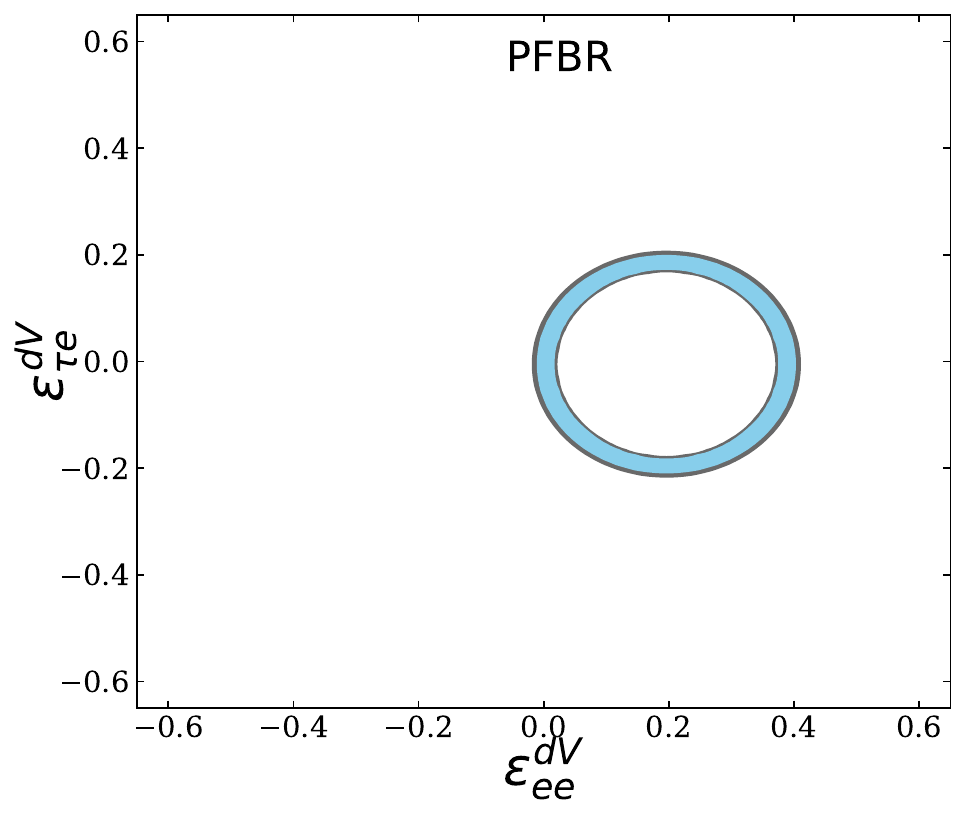}
\includegraphics[width=0.34\textwidth]{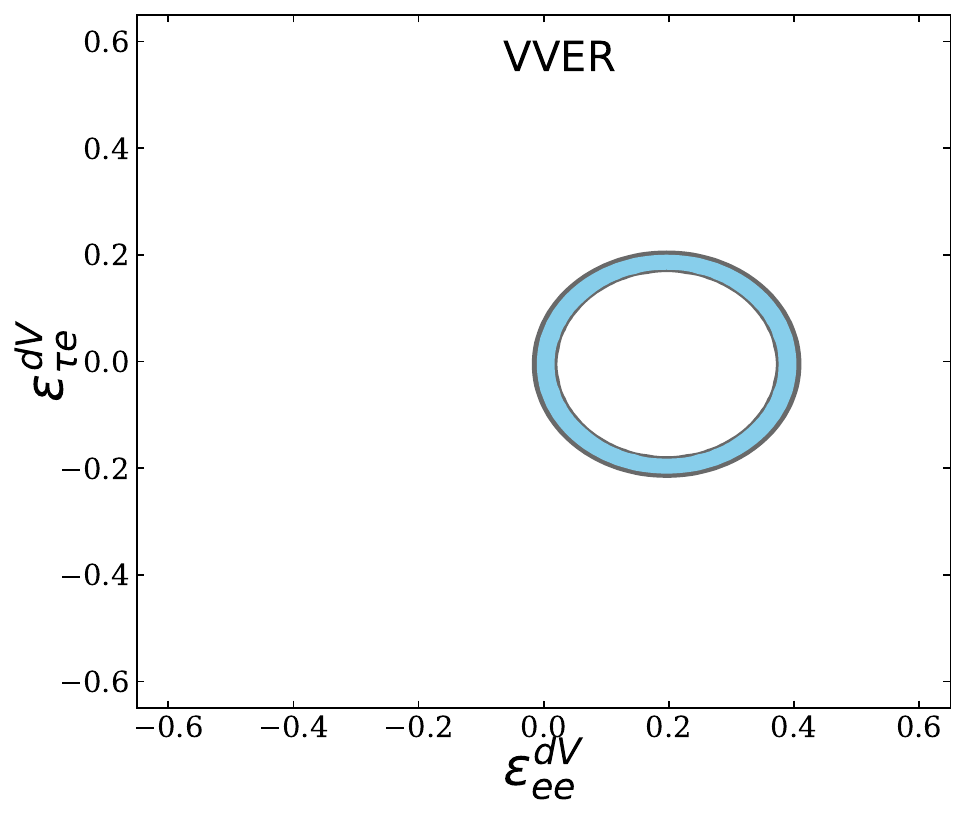}
\caption{ \label{fig:etau} The allowed regions of non-universal 
and flavor changing NSI parameters $\varepsilon_{ee}^{dV}$-$\varepsilon_{\tau e}^{dV}$
plane at 90$\%$ CL considering Apsara-U, Dhruva, PFBR and, VVER reactors as 
antineutrinos sources.} 
\end{figure*}
 reduced when background noise is present. As previously explained, enhancing 
 detector sensitivity can be achieved by utilizing targets made from combination of 
low and high Z materials. Figure~\ref{fig:tau_AlGe} shows a comparison of detector 
sensitivity when using targets made up of $Al_{2}O_{3}$ and $Ge$ each one has mass of 
5 kg.The results are compared with the same obtained from the CONUS+ 
experiment~\cite{CONUS:2024lnu}. It has been 
observed that a detector made up  from $Ge$ exhibits superior sensitivity compared to 
one made from $Al_{2}O_{3}$. Moreover, the orange shaded area represents 
allowed regions from combined analysis of results from both the detector that 
collectively boosts sensitivity while reducing parameter space degeneracy. This 
approach effectively optimizes detection capabilities, offering a clearer 
understanding of material-based sensitivity differences in detectors.
\begin{figure}[h]
\centering
\includegraphics[width=0.4\textwidth]{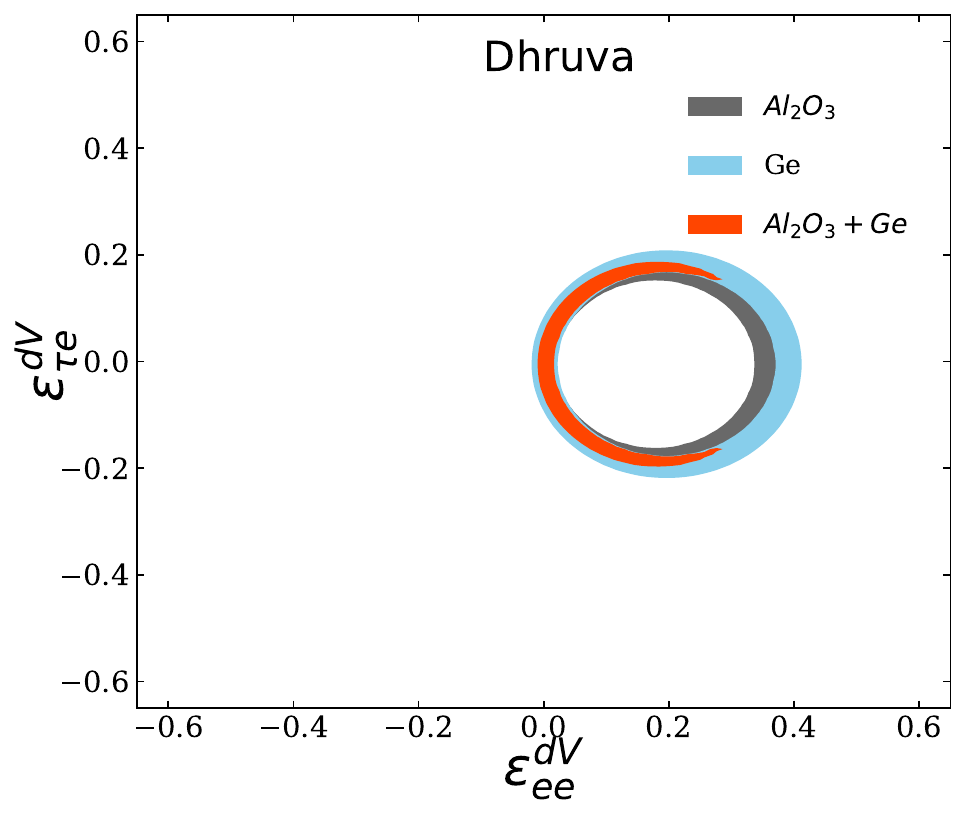}
\caption{ \label{fig:dedtau_AlGe} An expected  allowed regions of 
non-universal NSI parameters $\varepsilon_{ee}^{dV}$-$\varepsilon_{\tau e}^{dV}$
plane at 90$\%$ CL considering detector made up  with $Al_{2}O_{3}$ and Ge. 
Orange shaded region shows the combined results. Results are extracted 
considering a signal-to-background ratio of 1.0} 
\end{figure}
\begin{figure}[h]
\centering
\includegraphics[width=0.4\textwidth]{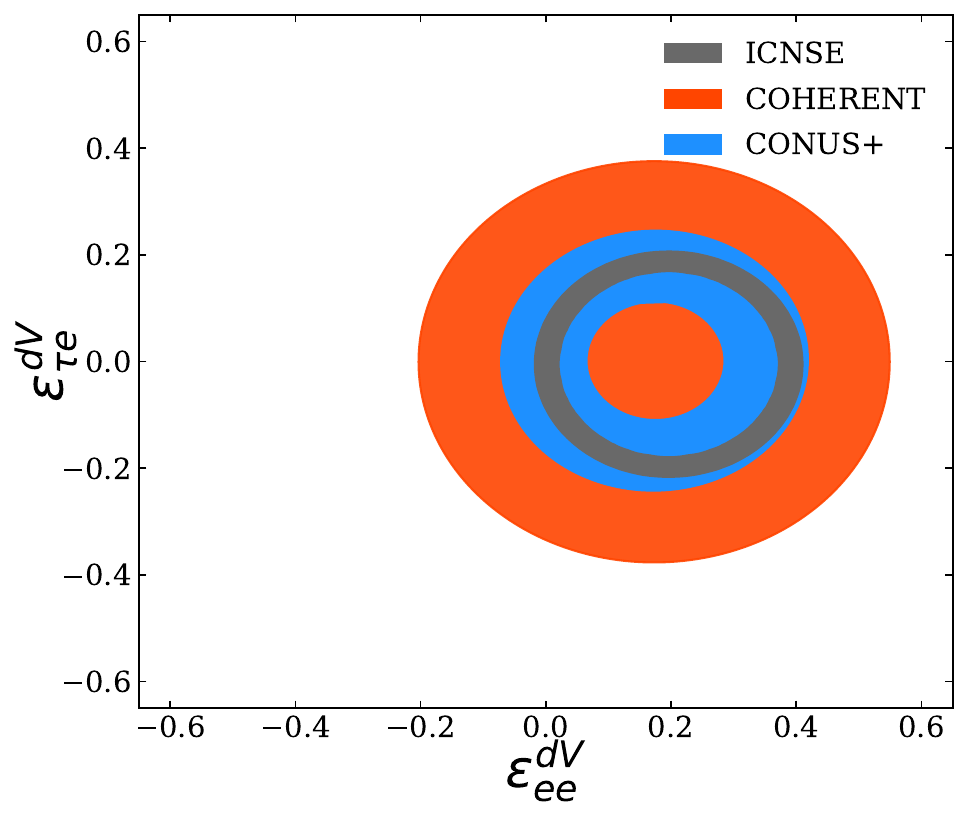}
\caption{ \label{fig:dedtau_global} Comparison of sensitivity
 of 10 kg $Al_{2}O_{3}$ detector at an exposure of one year  with other experimental 
observations in the $\varepsilon_{ee}^{dV}$-$\varepsilon_{\tau e}^{dV}$
plane at 90$\%$ CL The detector is placed at 10 m from 
Dhruva reactor core and with consideration
of expected signal-to-background ratio of one.} 
\end{figure}
\subsection{Sensitivity of the detector to both non-universal and flavor-changing 
NSI parameters}
The potential of a detector has been extracted considering non-universal and flavor 
changing NSI parameters one at a time, for NSI with d-type quark only. In this case 
$\varepsilon_{ee}^{dV}$ 
and $\varepsilon_{e\tau}^{dV}$ are considered to be non-zero for which \antinue s 
are affected (considering $\varepsilon_{ee}^{uV}$ = $\varepsilon_{e\tau}^{uV}$ = 0). The expected allowed region is between two 
circles as shown in Fig.~\ref{fig:etau} at 90 $\%$ CL 
considering sapphire detector without and with presence of background. The 
expected region can be understood by using the relation 
$\left (Zg_{p}^{V} + Ng_{n}^{V}\right )^{2}=  \left [Z\left(g_{p}^{V} 
+ \varepsilon_{ee}^{dV}\right) + N\left(g_{n}^{V} + 2\varepsilon_{ee}^{dV}\right 
)\right ]^{2} + \left [Z\varepsilon_{e \tau}^{dV} +2N\varepsilon_{e\tau}^{dV}\right 
]^{2}$.
 It shows the degeneracy in the NSI parameters.
 As previously stated, targets with distinct N/Z could be used to break
  the degeneracy. Figure~\ref{fig:dedtau_AlGe} shows the allowed region based on
  the combined analysis of results taking into consideration each 
  target of mass 5 kg of $Al_{2}O_{3}$ and $Ge$ in the existence of 
  background. Due to varying values of N and Z, it is found that 
  the circle's center and radius differ for both detectors. It is observed that 
  the combine result breaks the degeneracy, limits the allowed parameter space.
 Figure~\ref{fig:dedtau_global} shows the comparison on the sensitivity
  between expected results of the ICNSE sapphire detector, 
 the COHERENT CsI detector~\cite{Liao:2017uzy}, and CONUS+ HPGe 
detector~\cite{CONUS:2024lnu}. It shows that the ICNSE detector can limit more 
parameter space compared to CsI detector of COHERENT and CONUS+ experiment.
\begin{figure*}[ht]
\centering
\includegraphics[width=0.3\textwidth]{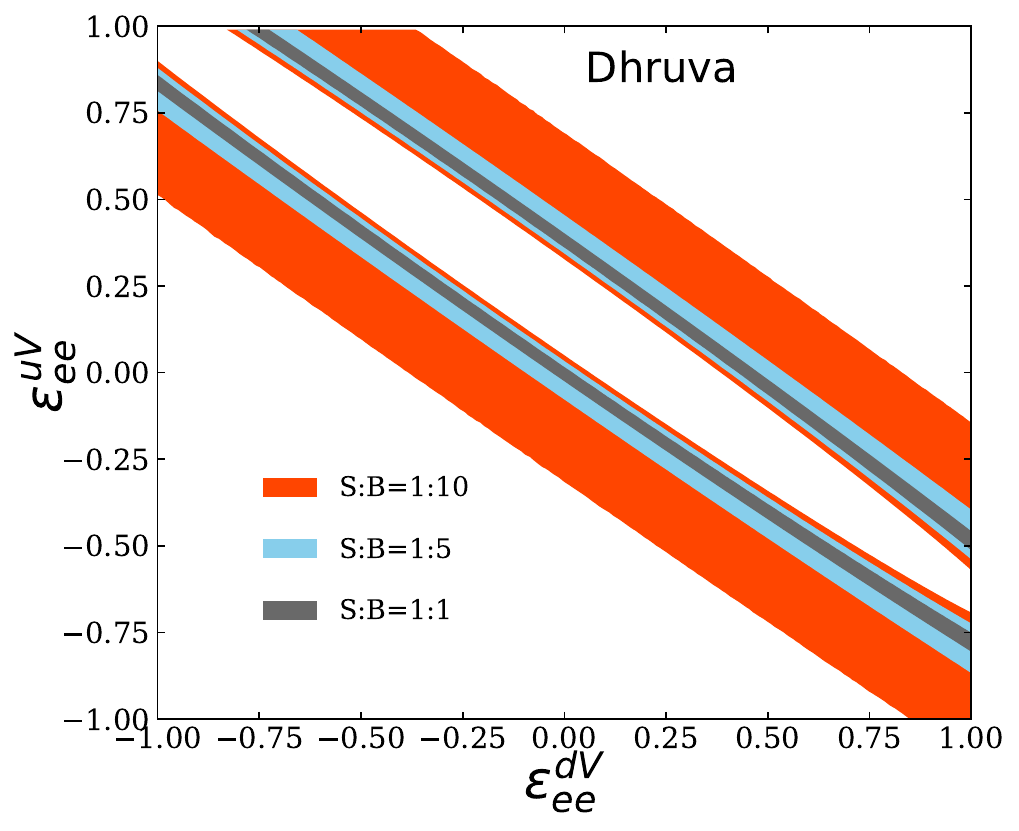}
\includegraphics[width=0.3\textwidth]{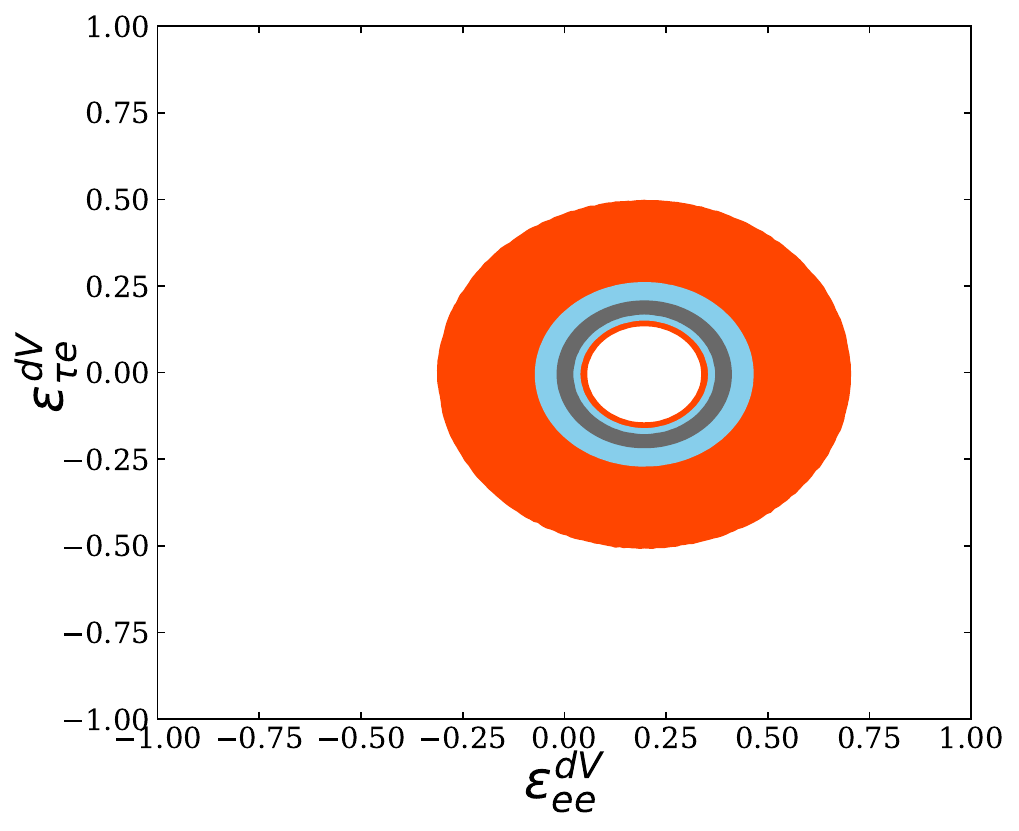}
\includegraphics[width=0.3\textwidth]{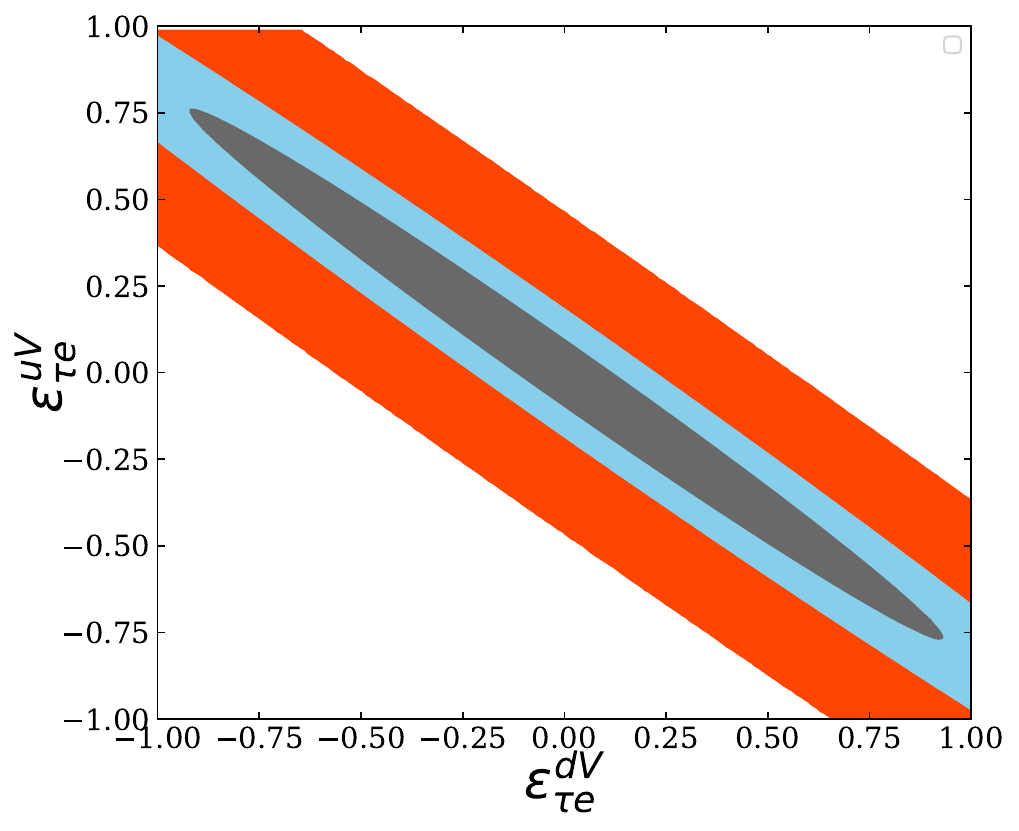}
\caption{ \label{fig:bkg_dhruva} Comparisons of sapphire detector sensitivity at 
90$\%$ CL for 
different values of signal-to-background ratio.} 
\end{figure*}
\begin{figure*}[ht]
\centering
\includegraphics[width=0.3\textwidth]{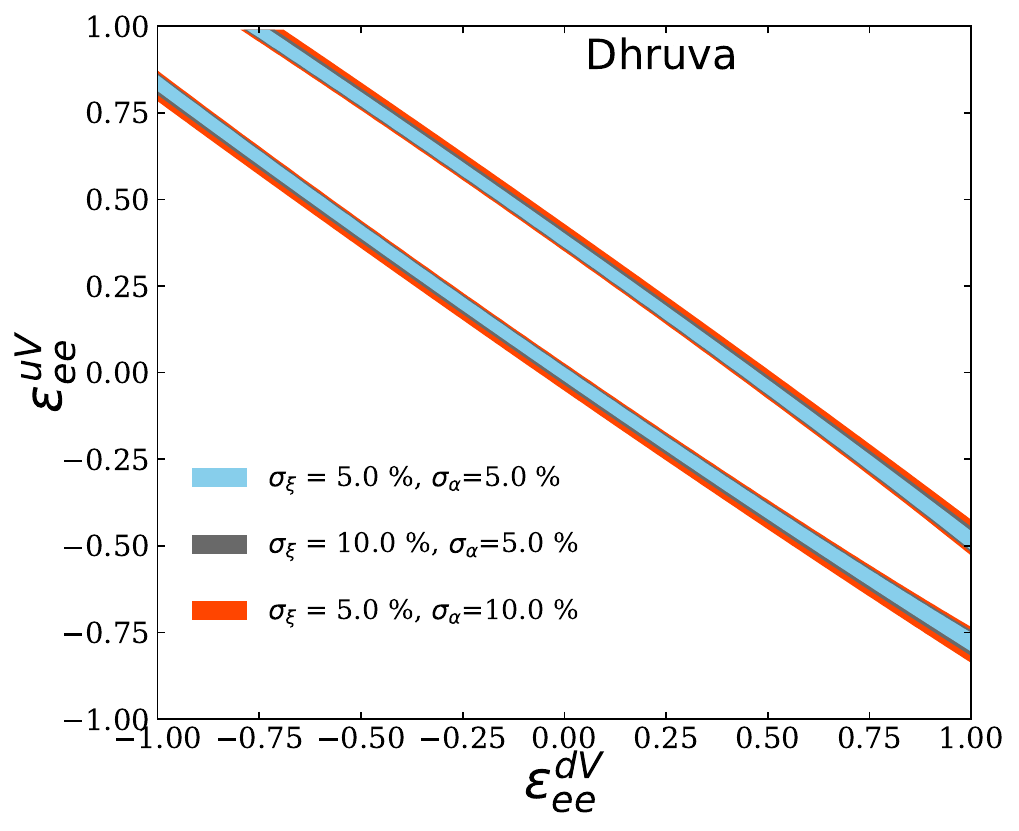}
\includegraphics[width=0.3\textwidth]{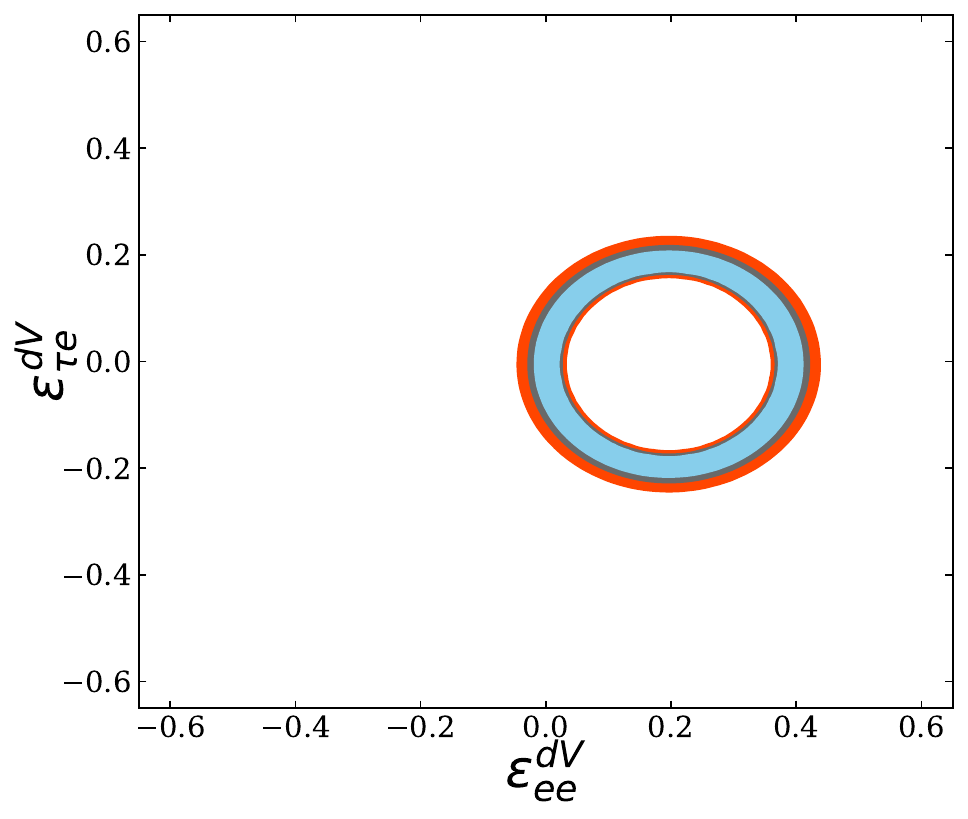}
\includegraphics[width=0.3\textwidth]{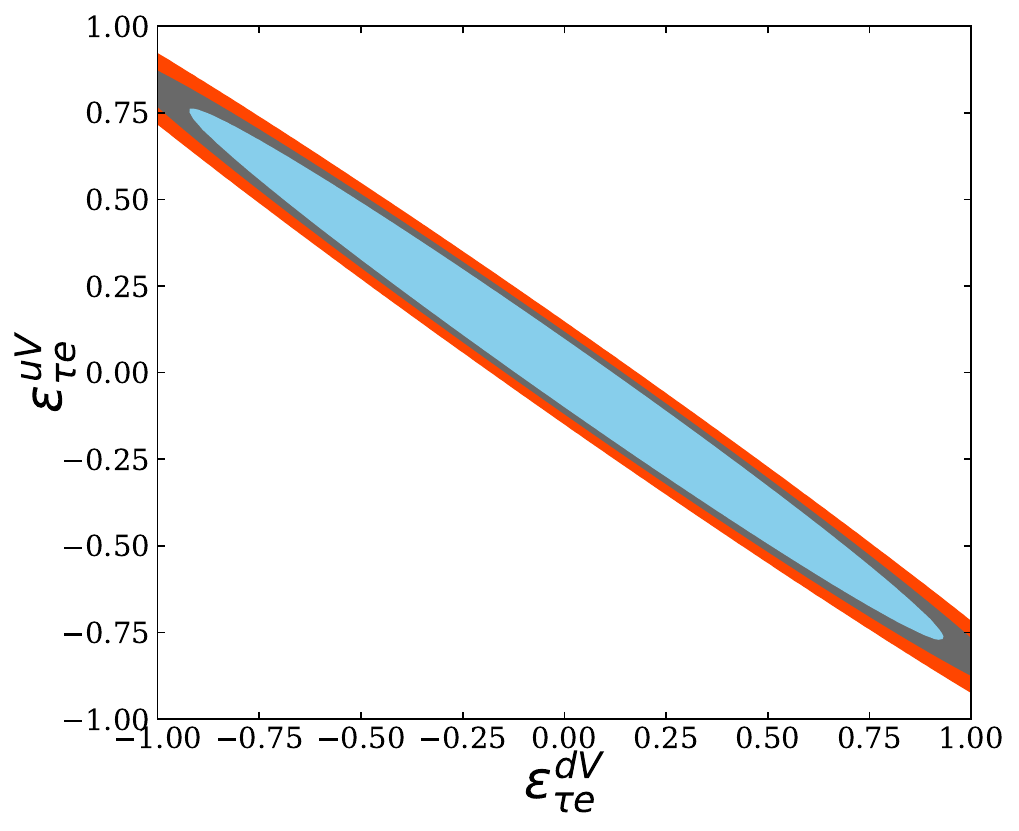}
\caption{ \label{fig:sys_dhruva} Comparisons of sapphire detector sensitivity at 
90$\%$ CL for different values of systematics. Results are extracted considering a 
signal-to-background ratio of one.} 
\end{figure*}
\begin{figure*}[ht]
\centering
\includegraphics[width=0.4\textwidth]{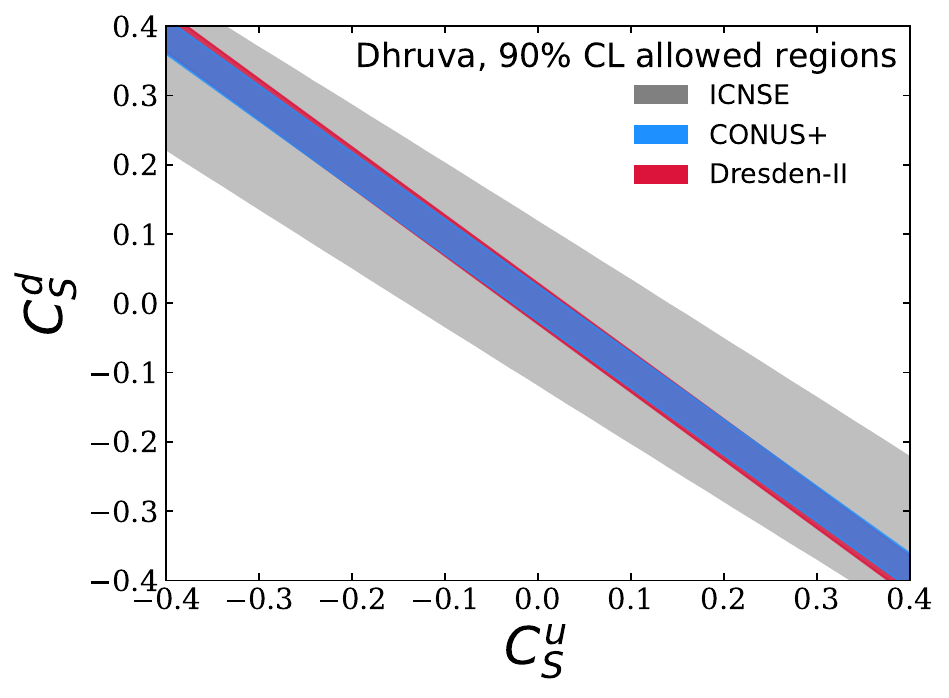}
\includegraphics[width=0.4\textwidth]{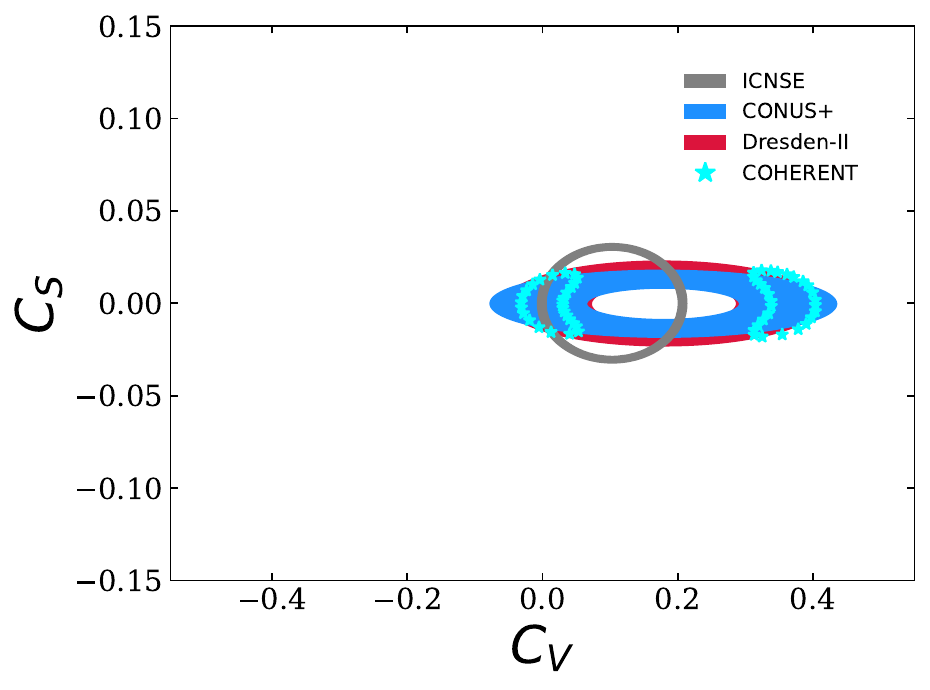}
\includegraphics[width=0.4\textwidth]{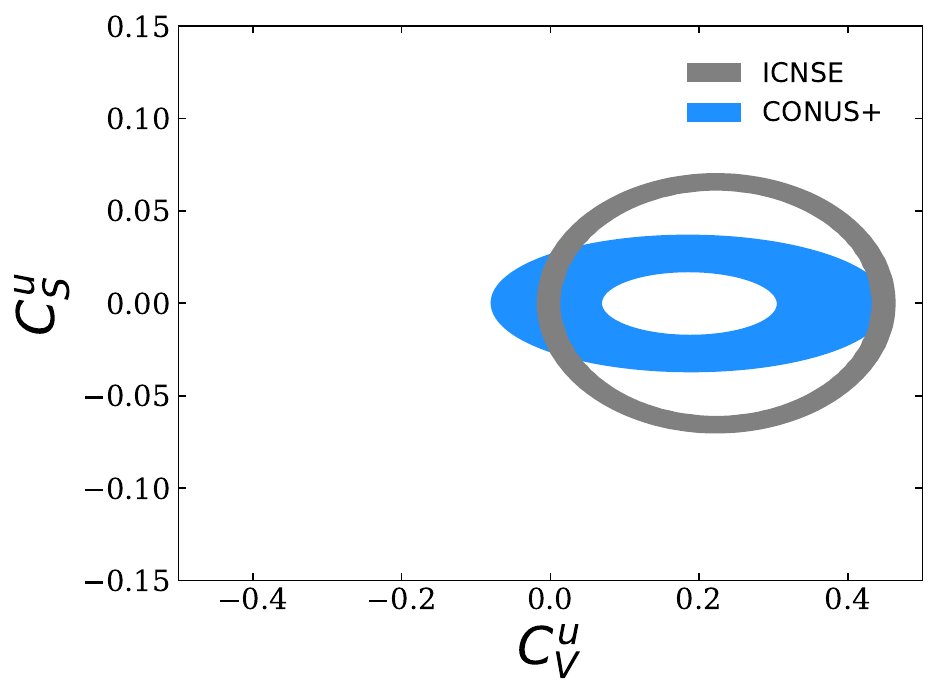}
\includegraphics[width=0.4\textwidth]{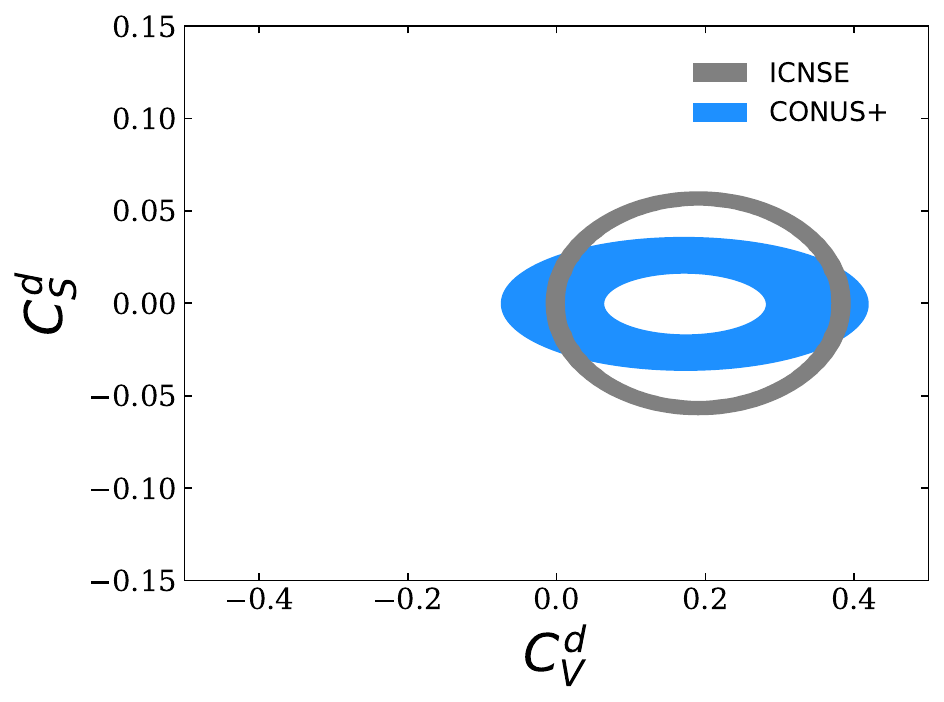}
\caption{ \label{fig:ngi}
The allowed regions on scalar and vector NGIs at 90$\%$ CL for a 
sapphire detector of mass 10 kg placed at 10 m from Dhruva reactor core.
It has been considered a signal-to-background ratio of one 
 and an exposure of one year.} 
\end{figure*}
\subsection{Effect of systematic and background on sensitivity}
Studies have been carried out to find the impact of background and systematic 
considering the Dhruva reactor as a source for $\antinue$s as there are several other 
basic science research are going on using neutron beam inside the reactor hall. 
Therefore, reactor related background is more at this site as compared to other 
facilities. In this case, it has been considered a sapphire detector of mass 10 kg
placed at distance of 10 m from the Dhruva reactor core for an exposure of one year.
It has been observed that the expected sensitivity of detector reduces with increase 
of background which is shown in Fig.~\ref{fig:bkg_dhruva}. At 90$\%$ CL, 
Fig.~\ref{fig:sys_dhruva} shows the detector sensitivity to various NSI parameters 
considering different combination of systematic values. It is found that with 
increasing the value of $\sigma_{\xi}$  to 10.0$\%$ from 5.0$\%$, the sensitivity of 
detector reduces. Further, for a given value of  $\sigma_{\xi}$, by increasing the 
$\sigma_{\alpha}$  to 10.0$\%$ from 5.0$\%$ the sensitivity of the detector has been 
reduced more as compared to earlier case. So, it is necessary to reduce the 
background and related systematic uncertainty. Similarly, studies are carried out to 
find the detector sensitivity at various signal-to-background ratio for given values 
of $\sigma_{\xi}$(5.0$\%$) and $\sigma_{\alpha}$(5.0$\%$). It can be noted here 
that further studies have been carried out considering a signal-to-background ratio 
of one and fixing the detector exposure.  
\subsection{Sensitivity to neutrino generalized interaction}                        %
The study has been carried out for an independent scalar interaction as
well as a combination of scalar and vector interactions and, an independent tensor interaction.
 The upper-left panel of Fig.~\ref{fig:ngi} shows
 the 90$\%$ CL allowed regions in the $C_{S}^{u}-C_{S}^{d}$ plane,
 focusing primarily on scalar interactions. The contour is a single band 
 due to minimal interference with the SM CE$\nu$NS cross-section. The upper right
 panel shows the allowed contour region at 90$\%$ CL while considering both scalar 
 and vector interactions. For simplicity, it has been assumed  universal quark 
couplings, $i.e.$, $C_{S}^{u}=C_{S}^{d} = C_{S}$ and $C_{V}^{u} = C_{V}^{d} = C_{V}$.
 The constrained region has a ring shape because of the interference effect from the 
 vector interaction. The lower-left and lower-right panels show results for scalar 
and vector interactions with u and d quarks only, respectively. As mentioned earlier, 
the NGIs involving d quarks are more constrained compared to those with u quarks, due 
to the neutron dominance in the CE$\nu$NS cross section. The results from the ICNSE 
are compared with the same obtained from Dresden-II~\cite{Colaresi:2022obx,Majumdar:2022nby}, 
COHERENT~\cite{COHERENT:2017ipa,Papoulias:2017qdn} and CONUS+\cite{CONUS:2024lnu,DeRomeri:2025csu} experiments.
\subsection{Sensitivity to non-universal NGI tensor parameters}
\begin{figure}
\centering
\includegraphics[width=0.4\textwidth,height=0.3\textwidth]{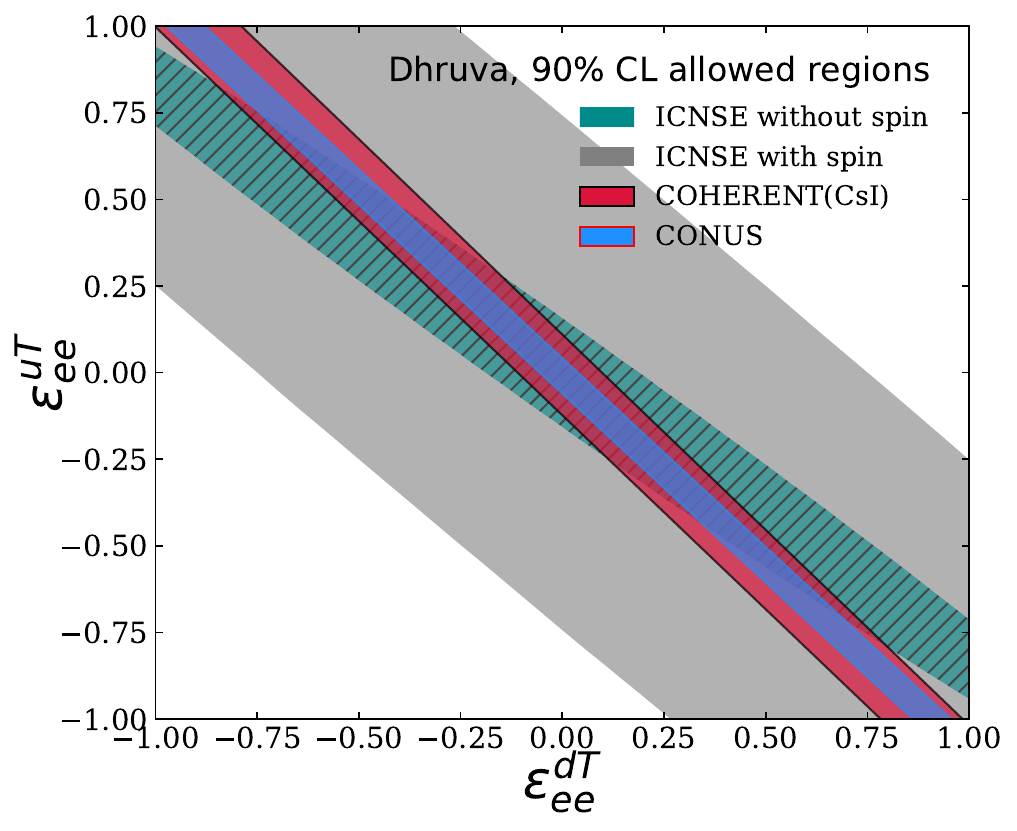}
\caption{ \label{fig:uTdT_AlGe} An expected  allowed regions of non-universal NSI 
tensor parameters $\varepsilon_{ee}^{dT}$-$\varepsilon_{ee}^{uT}$
plane at 90$\%$ CL considering detector made up  of with $Al_{2}O_{3}$. 
Result of the  ICNSE detector has been extracted considering a signal-to-background ratio of one.
Sensitivity at 90$\%$ CL due to COHERENT data has been taken form Ref.~\cite{Papoulias:2017qdn}.} 
\end{figure}

In this analysis, the Dhruva reactor is utilized as a source of electron 
 antineutrinos.
Figure~\ref{fig:uTdT_AlGe} illustrates the anticipated sensitivity to 
non-universal tensor NSI parameters in the $\varepsilon^{dT}_{ee}-\varepsilon^{uT}_{ee}$
 plane at a 90$\%$ CL considering both spin independent(dark cyan) and spin dependent(grey)
 neutrino generalized tensor interaction. The constraints on the NSI tensor parameters are derived 
 from the interactions of neutrinos with both u-type and d-type quarks. 
 The observed behavior of these results can be explained through the 
 arguments presented in the context of non-universal vector-type NSI interactions.
 It is observed that the sensitivity of sapphire detector reduces while considering the
 spin dependent neutrino generalized tensor interaction of $^{27}Al$.
  Additionally, the figure also shows the comparison of detector sensitivity with others results
  such as an expected sensitivity of the CONUS~\cite{CONUS:2021dwh} and the results
  obtained by analyzing the COHERENT data available in Ref.~\cite{Papoulias:2017qdn}.It has been found that
  sapphire detector can constrain most of the parameter space as like other experiments.
\section{SUMMARY}
\label{sec:summary}
In this article, we explore the expected sensitivity reach of the ICNSE detector
for constraining various parameters of nonstandard as well as generalized neutrino
interactions using the coherent elastic neutrino-nucleus scattering process. The
study has been performed considering an exposure of one year employing
$\overline{\nu}_e$s produced from a research reactor, which can be further employed
for measurement at other available power reactor. In the present study, high purity germanium 
and sapphire (Al$_2$O$_3$) detectors are considered.
It is found that the detector can reach for constraining most of the NSI as well as
NGI parameters space. The detector sensitivity is almost independent of reactor 
core. Further, there are degeneracies between different NSI parameters that appear 
while constraining more than one NSI parameter at a time using a single target. 
Results from a combination of two different types of detectors (Ge + 
Al$_2$O$_3$) having different neutron-to-proton ratios has one of the best 
sensitivities to the individual NSI parameters and can break the degeneracies between 
two NSI parameters for a large set of pairs of parameters.The expected sensitivity of 
the sapphire detector at a distance of 10 m from the reactor core are comparable with 
the other experiments and can put a stringent limit in the parameter space. The study 
also has been performed for with and without presence of background. The sensitivity 
deteriorates in presence of background noise. It has been observed that the detector 
has the potential for partially lifting these degeneracies when combining the results 
of two different detectors having different neutron to proton ratio. Results 
from the present study are compared with the same obtained from other experiments.
This study shows that the proposed neutrino experiment is sensitive to new 
physics of weak interaction and beyond. Finally, the sensitivity can be further 
improved by combining two different types of detectors with longer running time.
 \section*{ACKNOWLEDGMENTS}
The author would like to thank Dimitris Papoulias and Martin Hoferichter for the valuable suggestion 
and critical comments on the manuscripts. The author would also like to thank 
Abhijit Bhattacharyya and R. R. Sahu  for their technical suggestions.
\bibliography{NSI}
\bibliographystyle{JHEP}

\end{document}